\documentclass[twocolumn]{revtex4}
\usepackage{graphicx}  
\usepackage{dcolumn}   
\usepackage{bm}        
\usepackage{verbatim}   
\usepackage{graphicx}
\usepackage{footnote}
\usepackage{epstopdf}
\usepackage{psfrag}
\usepackage{epsfig}
\usepackage{placeins}
\usepackage{float}
\usepackage{subfigure}
\usepackage{amssymb}
\usepackage{amsmath}
\usepackage{lipsum} 
\usepackage[usenames, dvipsnames]{color}
\usepackage[colorlinks=true,linkcolor=blue,allcolors=blue]{hyperref}%

\begin{document}

\headsep = 40pt
\title{Nonreciprocal Electromagnetic Scattering from a Periodically Space-Time \\ Modulated Slab and Application to a Quasisonic Isolator}
\author{Sajjad Taravati, Nima Chamanara and Christophe Caloz}
\affiliation{Department of Electrical Engineering, Poly-Grames Research Center, Polytechnique Montr\'{e}al, Montr\'{e}al, QC, Canada}
\date{\today}

\begin{abstract}
Scattering of obliquely incident electromagnetic waves from periodically space-time modulated slabs is investigated. It is shown that such structures operate as nonreciprocal harmonic generators and spatial-frequency filters. For oblique incidences, low frequency harmonics are filtered out in the form of surface waves, while high-frequency harmonics are transmitted as space waves. In the quasi-sonic regime, where the velocity of the space-time modulation is close to the velocity of the electromagnetic waves in the background medium, the incident wave is strongly coupled to space-time harmonics in the forward direction, while in the backward direction it exhibits low coupling to other harmonics. This nonreciprocity is leveraged for the realization of an electromagnetic isolator in the quasi-sonic regime and is experimentally demonstrated at microwave frequencies.

\end{abstract}

\maketitle


\section{Introduction}

Space-time varying media, materials whose constitutive parameters are spatio-temporally modulated, were first studied in the context of traveling wave parametric amplifiers~\cite{Cullen_NAT_1958,Tien_JAP_1958,Oliner_Hessel_1959,Oliner_PIEEE_1963,Cassedy_PIEEE_1965,Cassedy_PIEEE_1967,Peng1969,Chu1969,Chu1972}.
In contrast to moving media, where the velocity of the medium is restricted to the speed of light, space-time modulated media offer both
subluminal and superluminal phase velocities. In contrast to static periodic media such as photonic crystals, periodic space-time media exhibit asymmetric, tilted dispersion~\cite{Oliner_PIEEE_1963,Cassedy_PIEEE_1965}. Moreover, superluminal space-time media produce electromagnetic bandgaps that are oriented vertically, compared to horizontal bandgaps in conventional photonic crystals and Bragg structures. These vertical bandgaps describe instabilities or unbounded growth~\cite{Cassedy_PIEEE_1965}. Harmonic generation is another feature of space-time media. In contrast to nonlinear harmonics, space-time harmonics are not governed by the classical Manley-Raw relations~\cite{Cassedy_PIEEE_1965}. This result stems from violation of energy conservation in space-time modulated media, as energy is pumped into the system through the modulation.

This topic has regained attention in the past years due to recently discovered exotic effects such as interband photonic transitions mediated by space-time varying media~\cite{Fan_PRB_1999} and associated nonreciprocity~\cite{Fan_NPH_2009,Fan_PRL_109_2012}, inverse Doppler effect in a shockwave induced photonic bandgap structure~\cite{reed2003reverseddoppler}, nonreciprocal space-time metasurfaces~\cite{Alu_PRB_2015,Shalaev_OME_2015,Fan_APL_2016}, and nonreciprocal antenna systems~\cite{Taravati_APS_2015,Alu_AWPL_2015,Alu_TAP_63_2015,Alu_PNAS_2016,Taravati_TAP_65_2017}. Nonreciprocity based on space-time modulation seems to offer a viable path towards integrated nonreciprocal photonic and electromagnetic devices. This technique addresses issues of conventional nonreciprocity techniques, such as incompatibility with integrated circuit technology in magnet-based nonreciprocity, signal power restrictions in nonlinear-based nonreciprocity~\cite{Fan_NP_2015}, and low power handling and frequency limitation in transistor-based nonreciprocity~\cite{carchon_2000,Taravati_NR07_2017}.

Previous research on space-time media has been mostly focused on propagation in infinite space-time media or normal incidence on a semi-infinite space-time modulated region. \emph{Oblique} electromagnetic incidence on a space-time modulated \emph{slab} has unique features that have been unexplored. This paper shows that such a structure operates as a nonreciprocal harmonic generator and filter. It is demonstrated that a space-time slab operates as a high-pass spatial frequency filter. For oblique incidence, low frequency harmonics are filtered out in the form of surface waves, while high frequency harmonics are transmitted as space waves. In the quasisonic regime, where the velocity of the space-time modulation is close to the velocity of the electromagnetic waves in the background medium, the incident wave is strongly coupled to space-time harmonics in the forward direction while in the backward direction it exhibits low coupling to other harmonics. This nonreciprocity is leveraged for the realization of an electromagnetic isolator in the quasisonic regime and is experimentally demonstrated at microwave frequencies.


The paper is organized as follows. Section~\ref{sec:gen_sol} presents an analytical solution for electromagnetic scattering from a space-time slab. Section~\ref{sec:sin_mod} derives analytical expressions describing electromagnetic scattering from a sinusoidally modulated space-time slab. Dispersion diagrams and isofrequency curves are described in Sec.~\ref{sec:nonr_trans}. Space-time transitions and their nonreciprocal nature are highlighted in details in Sec.~\ref{sec:nonr_trans}. Finally, Sec.~\ref{sec:isolator} presents an electromagnetic isolator based on non-reciprocal space-time transitions in a periodically modulated slab.
\section{GENERAL ANALYTICAL SOLUTION}\label{sec:gen_sol}
\begin{figure}
\begin{center}
\includegraphics[width=\columnwidth]{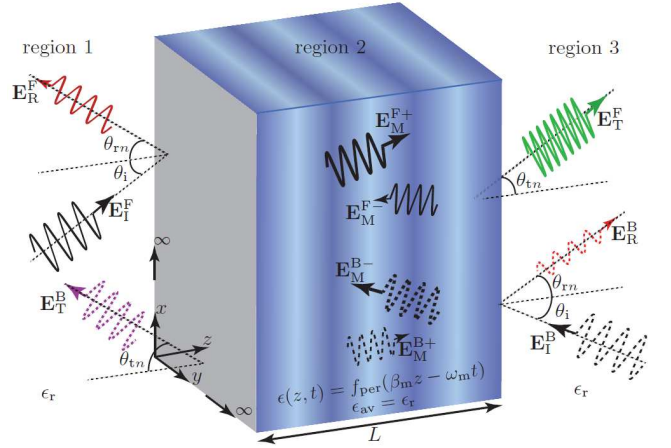}
\caption{Electromagnetic scattering from a periodically space-time modulated slab (region~2) sandwiched between two semi-infinite unmodulated media (regions~1 and~3). Due to the unidirectionality of the modulation, $\epsilon(z,t)=f_\text{per}(\beta_\text{m}z-\omega_\text{m}t)$, the system is nonreciprocal, with different temporal and spatial frequencies scattered in the two directions.}
\label{Fig:S-T_slab}
\end{center}
\end{figure}

The problem of interest is represented in Fig.~\ref{Fig:S-T_slab}. A plane wave, $\mathbf{E}_\text{I}$, impinges in the forward ($+z$) direction or backward ($-z$) direction under the angle $\theta_\text{i}$ on a periodically space-time modulated slab of thickness~$L$ sandwiched between two semi-infinite unmodulated media. Hereafter, the problem with the incident wave propagating towards the $+z$-direction, depicted at the top of Fig.~\ref{Fig:S-T_slab}, will be called the \emph{forward problem}, denoted by the superscript ``F'', while the problem with the incident wave propagating towards the $-z$-direction, depicted at the bottom of Fig.~\ref{Fig:S-T_slab}, will be called the \emph{backward problem}, denoted by the superscript ``B''. Note that, as illustrated in Fig.~\ref{Fig:S-T_slab}, the forward and backward problems both include forward and backward waves. The slab assumes the unidirectional forward relative permittivity
\begin{equation}
\epsilon(z,t)= f_\text{per}(\beta_\text{m}z-\omega_\text{m}t),
\label{eqa:Gen_perm}
\end{equation}
\noindent where $f_\text{per}(.)$ is an arbitrary periodic function of the space-time phase variable \mbox{$\xi=\beta_\text{m}z-\omega_\text{m}t$}, with $\beta _\text{m}$ being the spatial modulation frequency and $\omega_\text{m}$ the temporal modulation frequency. Taking the time derivative of a constant phase point in~\eqref{eqa:Gen_perm} yields $d\xi/dt=0=\beta_\text{m}dz/dt-\omega_\text{m}$,
\begin{equation}
v_\text{m}
=\frac{\omega_\text{m}}{\beta_\text{m}}.
\label{eqa:vm_omegam_betam}
\end{equation}
\noindent This velocity may be smaller or greater than the phase velocity of the background medium, which we define here as the velocity
\begin{equation}
v_\text{b}
=\frac{c}{\sqrt{\epsilon_\text{r}}},
\label{eqa:vr}
\end{equation}
\noindent where $c=1/\sqrt{\mu_0 \epsilon_\text{0}}$ is the speed of light in vacuum, and where $\epsilon_\text{r}$ is the relative permittivity common to media~1 and~3 and to the average permittivity of medium~2. The ratio between the modulation and background phase velocities,
\begin{equation}
\gamma
=\frac{v_\text{m}}{v_\text{b}},
\label{eqa:gamma}
\end{equation}
\noindent is called the \emph{space-time velocity ratio}. The limit $\gamma=0$ corresponds to a purely space-modulated medium, while the limit $\gamma=\infty$ corresponds to a purely time-modulated medium~\cite{Kalluri_2010}. Moreover, $\gamma=1$ corresponds to the space-time-modulated medium where the modulation propagates exactly at the same velocity as a wave in the background medium. We wish to calculate the fields scattered by the slab, namely the reflected fields, $\mathbf{E}_\text{R}^\text{F,B}$, the fields in the modulated medium, $\mathbf{E}_\text{M}^{\text{F,B};\pm}$, and the transmitted fields, $\mathbf{E}_\text{T}^\text{F,B}$, in Fig.~\ref{Fig:S-T_slab}.

Since the slab medium permittivity is periodic in space, with spatial frequency $\beta_\text{m}$, and in time, with temporal frequency $\omega_\text{m}$, it may be expanded in the space-time Fourier series
\begin{equation}
\epsilon (z,t) = \sum\limits_{k =  - \infty }^\infty  \tilde{\epsilon}_k {e^{-jk(\beta_\text{m}z-\omega_\text{m}t)}},
\label{eqa:Fourier_perm}
\end{equation}
where $\tilde{\epsilon}_k$ is the coefficient of the $k^\text{th}$ term and $\tilde{\epsilon}_0=\epsilon_\text{r}$. Moreover, assuming $\text{TM}_{y}$ or $E_y$ polarization, the electromagnetic fields inside the slab may be represented in the double space-time Bloch-Floquet form
\begin{subequations}\label{eqa:wave}
\begin{equation}
\mathbf{E}_\text{M}(x,z,t)=\mathbf{E}_\text{M}^+(x,z,t)+\mathbf{E}_\text{M}^-(x,z,t)=\sum_{n =  - \infty}^\infty  \left( {\mathbf{E}}_{n}^+ + {\mathbf{E}}_{n}^- \right),
\end{equation}
where the superscripts F and B have been omitted for notational simplicity and where the $\pm$ superscripts represent $\pm z$-propagating wave components. In~\eqref{eqa:wave},
\begin{equation}
{\mathbf{E}}_{n}^+ = \mathbf{\hat{y}} A_{n}^+ e^{-j ( k_x x+\beta_{0}^+ z-\omega_0 t)}     e^{  -j n(\beta_\text{m} z-\omega _\text{m} t )},
\label{eqa:E_forward}
\end{equation}
\begin{equation}
{\mathbf{E}}_{n}^- = \mathbf{\hat{y}} A_{n}^- e^{-j (k_x x -\beta_{0}^- z-\omega_0 t)}     e^{  -j n(\beta_\text{m} z-\omega _\text{m} t )}.
\label{eqa:E_backward}
\end{equation}
\end{subequations}
where $\beta_{0}$ and $\omega_0$  are the spatial and temporal frequencies of the fundamental temporal and spatial harmonics, respectively, in the slab, and $k_x= k_0 \sin(\theta_\text{i})=(\omega_0/v_\text{b}) \sin(\theta_\text{i})$ is the $x$-component of the spatial frequency, $\mathbf{k}$.

It is shown in Sec.~1 of~\cite{Taravati_PRB_SM_2016} that the Bloch-Floquet solution in~\eqref{eqa:wave} is valid everywhere except in the interval
\begin{equation}\label{eqa:sonic}
\gamma_\text{s,min}
=\sqrt{\frac{\epsilon_\text{r}}{\tilde{\epsilon}_0+\epsilon_\text{m}}}
\leq\gamma\leq
\sqrt{\frac{\epsilon_\text{r}}{\tilde{\epsilon}_0-\epsilon_\text{m}}}
=\gamma_\text{s,max},
\end{equation}
where $\tilde{\epsilon}_0$ is the average of $\epsilon(z,t)$, as seen in~\eqref{eqa:Fourier_perm}, and $\epsilon_\text{m}$ is the maximal (symmetric) variation of $\epsilon(z,t)$ from $\tilde{\epsilon}_0$, and is called the modulation depth. Upon multiplication by $v_\text{b}$ and usage of~\eqref{eqa:gamma} and~\eqref{eqa:vr}, this interval may also be expressed in terms of the modulation velocity as
\begin{equation}\label{eqa:sonic_vm}
v_\text{m,s,min}
=\frac{c}{\sqrt{\tilde{\epsilon}_0+\epsilon_\text{m}}}
\leq v_\text{m}\leq
\frac{c}{\sqrt{\tilde{\epsilon}_0-\epsilon_\text{m}}}
=v_\text{m,s,max},
\end{equation}
\noindent and is called the ``sonic regime''~\cite{Oliner_Hessel_1959} in analogy with a similar interval first identified in acoustic space-time modulated problems. It has been established that, in the case of the (nonperiodic) space-time slab $\epsilon(z,t)=\text{rect}(\beta_\text{m}z-\omega_\text{m}t)$, the sonic regime~\eqref{eqa:sonic_vm} supports both a space-like reflected wave and a time-like reflected wave, whereas only a space-like reflected wave exists when $v_\text{m}<v_\text{m,s,min}$ and only a time-like reflected wave exists when $v_\text{m}>v_\text{m,s,max}$~\cite{Biancalana_PRE_2007}. The sonic interval thus represents a regime requiring a special mathematical treatment, that has not yet been reported in the literature to the best of the authors' knowledge. When the space-time modulation is made periodic, as in~\eqref{eqa:Fourier_perm}, the same phenomenon occurs for each interface, and therefore the interval~\eqref{eqa:sonic_vm} still corresponds to the same sonic regime. In the middle of the sonic interval, i.e. at $\gamma=1$ or $v_\text{m}=v_\text{b}$, all the forward space-time harmonics merge into a single dispersion curve, as will be explained later, leading to a shock wave as in the phenomenon of sound barrier breaking in acoustics.

To find the unknown coefficients, $\beta_{0}^+$, $\beta_{0}^-$, $A_{n}^+$ and $A_{n}^-$ in~\eqref{eqa:wave}, we shall first fix $\omega_0$, as the source frequency, and then find the corresponding discrete $\beta_{0}$ solutions, $\beta_{0p}$, forming the dispersion diagram of the slab. Next, we shall apply the spatial boundary conditions at the edges of the slab, i.e. at $z=0$ and $z=L$, for all the $(\omega_0,\beta_{0p})$ states in the dispersion diagram, which will provide the unknown slab coefficients $A_{n(p)}^+$ and $A_{n(p)}^-$ in~\eqref{eqa:E_forward} and~\eqref{eqa:E_backward}, respectively, and the corresponding coefficients in the unmodulated regions, i.e. the fields everywhere.

The source-less wave equation reads
\begin{equation}
\nabla^2 \mathbf{E}_\text{M}(x,z,t) - \frac{1}{{{c^2}}}\frac{{{\partial ^2} \left[\epsilon (z,t)\mathbf{E}_\text{M}(x,z,t) \right]}}{{\partial {t^2}}}=0.
\label{eqa:wave_eq}
\end{equation}
\noindent Inserting~\eqref{eqa:wave} into the first term of~\eqref{eqa:wave_eq}, and~\eqref{eqa:A-product_E_eps} in Sec.~2 of~\cite{Taravati_PRB_SM_2016} [product of~\eqref{eqa:Fourier_perm} and~\eqref{eqa:wave}] into the second term of~\eqref{eqa:wave_eq}, and next using~\eqref{eqa:A-summation_A}, yields the relation
\begin{equation}
A_{n}^\pm \left[ \frac{ k_x^2+ (\beta_{0} \pm n \beta_\text{m})^2  }{\left[(\omega_0+n\omega_\text{m})/c \right]^2 }\right]
-  \sum\limits_{k =  - \infty }^\infty  \tilde{\epsilon}_k A_{n-k}^\pm  =0.
\label{eqa:recurs_gen}
\end{equation}
\noindent Equation~\eqref{eqa:recurs_gen} may be cast, after truncation to $2N+1$ terms, to the matrix form
\begin{equation}
[K^\pm]\cdot[A^\pm]=0,
\label{eqa:matrix_eq}
\end{equation}
where $[K^\pm]$ is the $(2N+1)\times(2N+1)$ matrix with elements
\begin{equation}
\begin{split}
K_{nn}^\pm &= \left[ \frac{ k_x^2+ (\beta_{0} \pm n \beta_\text{m})^2  }{\left[(\omega_0+n\omega_\text{m})/c \right]^2 }\right]-  \epsilon_0,   \\
K_{nk}^\pm &= -  \tilde{\epsilon}_{n-k},\quad\text{for }n\neq k,
\end{split}
\label{eqa:K_matrix}
\end{equation}
and $[A^\pm]$ is the $(2N+1)\times 1$ vector containing the $A_n^\pm$ coefficients. The dispersion relation is then given by
\begin{equation}
\text{det}\left\{[K^\pm]\right\}=0,
\label{eqa:det_gen}
\end{equation}
and the $(2N+1)$ forward and backward dispersion curves $\beta_{0p}(\omega_0)$, whose number is here finite due to truncation but theoretically infinite, are formed by solving this equation separately for the $\pm z$-propagating waves for a given set of modulation parameters $\omega_\text{m}$, $\beta_\text{m}$ and $\tilde{\epsilon}_k$, and for values of $\omega_0$ swept across the temporal frequency range of interest. Note that each point $(\beta_{0p},\omega_0)$ represents a \emph{mode} of the medium, itself constituted of an infinite number of oblique \emph{space-time harmonics} corresponding to modes at other frequencies, since such a point is a solution to the complete wave equation by virtue of~\eqref{eqa:det_gen}.

Once the dispersion diagram has been constructed, i.e. once the $\beta_{0p}^\pm$ states, solutions to~\eqref{eqa:wave_eq}, have been determined versus $\omega_0$, the unknown field amplitudes $A_{np}^\pm$ in the slab are found by solving~\eqref{eqa:matrix_eq} after determining the $A_{0p}^\pm$ terms satisfying boundary conditions. These terms are derived in Sec.~3 of~\cite{Taravati_PRB_SM_2016} as
\begin{subequations}\label{eqa:T_forward_backward}
\begin{equation}\label{eqa:T_forward}
A_{0p}^{\text{F}+}=\frac{E_\text{0}  k_\text{0} [\cos(\theta_\text{i}^+)+\cos(\theta_{\text{r}0}^+) ] }{  \beta_{0p}^+ +k_0 \cos(\theta_{\text{r}0}^+) - \frac{\beta_{0p}^- -k_0 \cos(\theta_{\text{r}0}^+)}{e^{j (\beta_{0p}^+ + \beta_{0p}^-) L}}  \frac{\beta_{0p}^+ -k_0 \cos(\theta_{\text{t}0}^+) }{\beta_{0p}^- + k_0 \cos(\theta_{\text{t}0}^+)} },
\end{equation}\begin{equation}\label{eqa:R_forward}
A_{0p}^{\text{F}-}=A_{0p}^{\text{F}+} e^{-j (\beta_{0p}^+ + \beta_{0p}^-) L}  \frac{ \beta_{0p}^+ -k_{0} \cos(\theta_{\text{t}0}^+) }{\beta_{0p}^-  +k_{0} \cos(\theta_{\text{t}0}^+) },
\end{equation}
\end{subequations}
for the forward problem, and
\begin{subequations}\label{eqa:T_forward_backward_b}
\begin{equation}\label{eqa:T_backward}
A_{0p}^{\text{B}-}=\frac{E_\text{0}  k_\text{0} [\cos(\theta_\text{i}^-)+\cos(\theta_{\text{r}0}^-) ] e^{j k_0 \cos(\theta_{\text{i}}^-) L} }  { \frac{\beta_{0p}^- +k_0 \cos(\theta_{\text{r}0}^-) }{e^{-j \beta_{0p}^- L}}  - \frac{\beta_{0p}^+ -k_0 \cos(\theta_{\text{r}0}^-)}{e^{j \beta_{0p}^+  L}}  \frac{\beta_{0p}^- -k_0 \cos(\theta_{\text{t}0}^-) }{\beta_{0p}^+ + k_0 \cos(\theta_{\text{t}0}^-)} },
\end{equation}
\begin{equation}\label{eqa:R_backward}
A_{0p}^{\text{B}+}=A_{0p}^{\text{B}-}  \frac{ \beta_{0p}^- -k_{0} \cos(\theta_{\text{t}0}^-) }{\beta_{0p}^+  +k_{0} \cos(\theta_{\text{t}0}^-) },
\end{equation}
\end{subequations}
\noindent for the backward problem, where $k_0=\omega_0\sqrt{\epsilon_\text{r}}/c$ is the spatial frequency in the unmodulated media. As expected from the unidirectionality of the perturbation [Eq.~\eqref{eqa:Gen_perm}], we have $A_{0p}^{\text{F}+}\neq A_{0p}^{\text{B}-}$ and $A_{0p}^{\text{F}-}\neq A_{0p}^{\text{B}+}$. It may be easily verified that in the particular case where the temporal perturbation is switched off ($\omega_\text{m}=0$), so that $\beta_{0,p}^+=\beta_{0,p}^-=\beta_{0,p}$, these inequalities transform to equalities after compensating for the round-trip phase shift $-2\beta_{0,p}L$, as expected for the resulting reciprocal system.

From this point, the scattered fields in the unmodulated media, also derived in Sec.~3 of~\cite{Taravati_PRB_SM_2016}, are found as
\begin{subequations}\label{eqa:E_RT_forward}
\begin{equation}\label{eqa:E_RF}
\begin{split}
\mathbf{E}_\text{R}^\text{F} = \mathbf{\hat{y}} \sum\limits_{n =  - \infty }^\infty & e^{-j \left[ k_{0} \sin(\theta_{\text{i}}) x- k_{0n} \cos(\theta_{\text{r}n}) z  -(\omega _0 +n\omega _\text{m}) t \right] } \\
& \cdot \left[  \sum_{p =  - \infty}^\infty  \left( A_{np}^{\text{F}+}+ A_{np}^{\text{F}-}  \right) - E_0^+ \delta _{n0}\right],
\end{split}
\end{equation}
\begin{equation}\label{eqa:E_TF}
\begin{split}
\mathbf{E}_\text{T}^\text{F}= & \mathbf{\hat{y}} \sum\limits_{n =  - \infty }^\infty   e^{-j \left[ k_{0} \sin(\theta_\text{i}) x+ k_{0n} \cos(\theta_{\text{t}n}) z-(\omega _0 +n\omega_\text{m}) t  \right] }, \\
 . & \sum_{p =  - \infty}^\infty \left( A_{np}^{\text{F}+} e^{-j (\beta_{0p} + n \beta_\text{m})L} +A_{np}^{\text{F}-} e^{j (\beta_{0p} - n \beta_\text{m})L} \right).
\end{split}
\end{equation}
\end{subequations}
and
\begin{subequations}\label{eqa:E_RT_backward}
\begin{equation}\label{eqa:E_RB}
\begin{split}
\mathbf{E}_\text{R}^\text{B} = \mathbf{\hat{y}} \sum\limits_{n =  - \infty }^\infty   e^{-j \left[ k_{0} \sin(\theta_{\text{i}}) x+ k_{0n} \cos(\theta_{\text{r}n}) z -(\omega _0 +n\omega _\text{m}) t \right] } \\ \cdot \bigg[\sum_{p =  - \infty}^\infty \big( A_{np}^{\text{B}+} e^{-j (\beta_{0p} + n \beta_\text{m})L}  + A_{np}^{\text{B}-} e^{j (\beta_{0p} - n \beta_\text{m})L}\big)\\
\qquad \qquad \qquad \qquad \qquad \qquad -E_0^- {\delta _{n0}} e^{j k_\text{0} L}\bigg],
\end{split}
\end{equation}
\begin{equation}\label{eqa:E_TB}
\begin{split}
\mathbf{E}_\text{T}^\text{B} =& \mathbf{\hat{y}} \sum\limits_{n =  - \infty }^\infty   \left(  A_{np}^{\text{B}+} +A_{np}^{\text{B}-}\right)  \\& \cdot e^{-j \left[  k_{0} \sin(\theta_\text{i}) x- k_{0n} \cos(\theta_{\text{t}n}) z -(\omega _0 +n\omega_\text{m}) t \right] },
\end{split}
\end{equation}
\end{subequations}
where $k_{0n}=(\omega_0+n\omega_\text{m}/v_\text{b})$.

\begin{figure}
\begin{center}
\includegraphics[width=1.05\columnwidth]{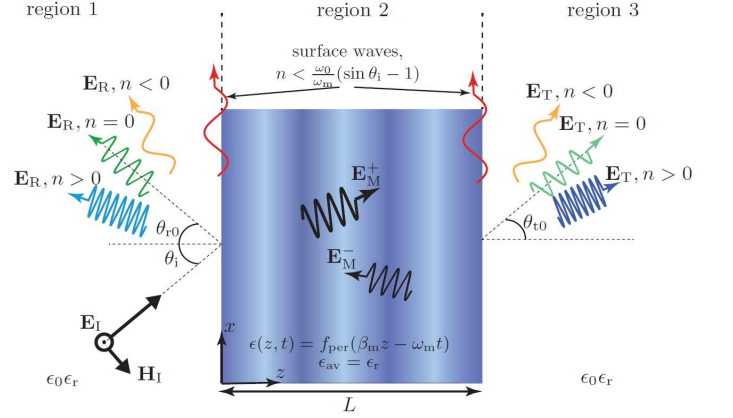}
\caption{Scattered space-time harmonics (shown here for the forward problem).}
\label{Fig:S-T_slab_2D}
\end{center}
\end{figure}

The scattering angles of the different space-time harmonics in~\eqref{eqa:E_RT_forward} for the forward problem, are represented in Fig.~\ref{Fig:S-T_slab_2D}. They are obtained from the corresponding Helmholtz relations
\begin{subequations}\label{eqa:scat_angl}
\begin{equation}\label{eqa:refl_angl}
[k_{0} \sin(\theta_{\text{i}})]^2+[k_{0n} \cos(\theta_{\text{r}n})]^2=k_{0n}^2,
\end{equation}
and
\begin{equation}\label{eqa:trans_angl}
[k_{0} \sin(\theta_{\text{i}})]^2+[k_{0n} \cos(\theta_{\text{t}n})]^2=k_{0n}^2,
\end{equation}
yielding
\begin{equation}\label{eqa:refl_trans_angl}
\sin(\theta_{\text{r}n})=\sin(\theta_{\text{t}n})=\frac{\sin(\theta_\text{i})}{1+n\omega _\text{m}/\omega_0},
\end{equation}
\end{subequations}
\noindent where $\theta_{\text{r}n}$ and $\theta_{\text{t}n}$ are the reflection and transmission angles of the $n^\text{th}$ space-time harmonic. Equation~\eqref{eqa:refl_trans_angl} describes the space-time spectral decomposition of the scattered wave. The reflection and transmission angles for a given harmonic $n$ are equal, due to phase matching, i.e due to the unique tangential wavenumber, $k_x=k_0\sin(\theta_\text{i})$ in all the regions [Eqs.~\eqref{eqa:refl_angl} and~\eqref{eqa:trans_angl}]. The harmonics in the $n$-interval $[\omega_0(\sin\theta_\text{i}-1)/\omega_\text{m},+\infty[$ are scattered (reflected and transmitted) at angles ranging from $\pi/2$ to $0$ through $\theta_\text{i}$ for $n=0$. The harmonics outside of this interval correspond to imaginary $k_{znp}^\pm$ and are hence not scattered. Rather, they travel as surface waves along the boundary. In the modulated medium, the scattering angles are found from the dispersion relation as
\begin{equation}\label{eqa:mod_angl}
\tan(\theta_{np}^\pm)=\frac{k_x}{k_{znp}^\pm}=\frac{k_0 \sin(\theta_{\text{i}})}{\beta_{0p}^\pm \pm n \beta_\text{m}}.
\end{equation}

Wave scattering in general periodic space-time modulated media share some similarities with stimulated Brillouin scattering, a nonlinear process whereby light interacts coherently with externally-applied acoustic variations in a medium and energy can be transferred back and forth between them~\cite{Brillouin_1922,Boyd_2003}. As the result of the interaction, frequency and wavenumber of a fraction of the transmitted light wave changes, in the same way if it were diffracted by an oscillating and moving grating. The phenomenon behind the stimulated Brillouin scattering is electrostriction, where frequency and momentum of a photon will change through a scattering process that releases a phonon~\cite{Fan_JLWT_2011,Fan_OE_2012}. Electrostriction may be represented as the tendency of materials to become compressed in the presence of an electric field.

\section{SINUSOIDALLY MODULATED SLAB}\label{sec:sin_mod}
We next consider a sinusoidal forward space-time permittivity as a particular case of the general periodic permittivity in~\eqref{eqa:Gen_perm}, namely
\begin{equation}
\epsilon(z,t)= \epsilon_\text{r}+\epsilon_\text{m} ~\cos(\beta_\text{m}z-\omega_\text{m}t).
\label{eqa:sin_perm}
\end{equation}
Such a permittivity has been used in~\cite{Cullen_NAT_1958} for the realization of a traveling-wave parametric amplifier. For the computation of the solution derived in Sec.~\ref{sec:gen_sol}, we write the expression~\eqref{eqa:sin_perm} in terms of its space-time Fourier components, i.e.
\begin{subequations}\label{eqa:eps_sin_exp}
\begin{equation}
\epsilon (z,t) = {\tilde{\epsilon}_{- 1}}{e^{ -j(\beta_\text{m}z-{\omega _{\text{m}}}t )}} + \tilde{\epsilon}_0 + \tilde{\epsilon}_{+1}e^{ + j({\beta _{\text{m}}}z-{\omega _{\text{m}}}t )},
\end{equation}
with
\begin{equation}
\tilde{\epsilon}_{-1}=\tilde{\epsilon}_{+1}= \epsilon_\text{m}/2\quad\text{and}\quad\tilde{\epsilon}_0=\epsilon_\text{r}.
\end{equation}\end{subequations}
Inserting~\eqref{eqa:eps_sin_exp} into~\eqref{eqa:recurs_gen} and subsequently following~\cite{Taravati_TAP_65_2017}, we find the analytic expressions
\begin{equation}
A_{np}^\pm = A_{n+1,p}^\pm \frac{1}{ - K_{np}^\pm + \frac{1}{K_{n-1,p}^\pm+ \frac{1}{ - K_{n-2,p}^\pm + \frac{1}{K_{n-3,p}^\pm+\ldots }} }}
\label{eqa:A_neg}
\end{equation}
for $n<0$, and
\begin{equation}
A_{np}^\pm = A_{n-1,p}^\pm \frac{1}{ - K_{np}^\pm + \frac{1}{K_{n+1,p}^\pm+ \frac{1}{ - K_{n+2,p}^\pm + \frac{1}{K_{n+3,p}^\pm+\ldots }} }},
\label{eqa:A_pos}
\end{equation}
for $n>0$, where
\begin{equation}
K_{np}^\pm=\frac{2 \epsilon_\text{r}}{\epsilon_\text{m}}  \left[1- \left( \frac{ k_x^2+ (\beta_{0}^\pm \pm n \beta_\text{m})^2  }{\left[(\omega_0+n\omega_\text{m})/v_\text{b} \right]^2 }\right)\delta_{nn} \right].
\label{eqa:K_np}
\end{equation}
The sonic interval associated with the sinusoidal permittivity in~\eqref{eqa:sin_perm} is obtained by inserting~$\tilde{\epsilon}_0$ into~\eqref{eqa:sonic} as
\begin{equation}\label{eqa:sonic_sin}
\gamma_\text{s,min}
=\frac{1}{\sqrt{1+\epsilon_\text{m}/\epsilon_\text{r}}}  \leq \gamma \leq \frac{1}{\sqrt{1-\epsilon_\text{m}/\epsilon_\text{r}}}
=\gamma_\text{s,max},
\end{equation}
where it is understood that $|\epsilon_\text{m} \cos(\beta_\text{m}z-\omega_\text{m}t)| \leq \epsilon_\text{m}$. Following again~\cite{Taravati_TAP_65_2017}, we also find the following analytic form for the dispersion relation of the slab:
\begin{equation}
\begin{split}
 & \frac{1}{ - K_{p,-1}^\pm + \frac{1}{K_{p,-2}^\pm+ \frac{1}{ - K_{p,-3}^\pm + \frac{1}{K_{p,-4}^\pm+\ldots }} }}+K_{0p}^\pm \\
  & \qquad \qquad + \frac{1}{ - K_{p,1}^\pm + \frac{1}{K_{p,2}^\pm+ \frac{1}{ - K_{p,3}^\pm + \frac{1}{K_{p,4}^\pm+\ldots }} }}=0.
  \end{split}
 \label{eqa:dispers_eq}
\end{equation}
This equation, which uses the $K_{np}^\pm$'s in~\eqref{eqa:K_np}, provides, for a given set
of modulation parameters ($\epsilon_\text{m}$, $\epsilon_\text{r}$, $\omega_\text{m}$, $\beta_\text{m}$, $\gamma$) and variable
$\omega_0$, the periodic dispersion diagram ($\beta_{0p}$'s) of the system.

Finally, the local space-time phase velocity and characteristic impedance read
\begin{subequations}\label{eqa:sin_velocity_imp}
\begin{equation}
v(z,t)=\frac{c}{\sqrt{\epsilon(z,t)}}
= \frac{c}{\sqrt{\epsilon_\text{r}+\epsilon_\text{m}\cos(\beta_\text{m}z-\omega_\text{m}t)} }
\label{eqa:sin_velocity}
\end{equation}
and
\begin{equation}
Z_0(z,t)= \sqrt{\frac{\mu}{\epsilon(z,t)}}=\sqrt{\frac{\mu_0}{\epsilon_0 \left(\epsilon_\text{r}+\epsilon_\text{m} ~\cos(\beta_\text{m}z-\omega_\text{m}t) \right)}},
\label{eqa:sin_imped}
\end{equation}
\end{subequations}
respectively. Equation~\eqref{eqa:sin_velocity_imp} indicate that the scattering angles and matching level both vary in space and time when $\epsilon_\text{m}\neq 0$.
\section{QUASI-SONIC NONRECIPROCITY}\label{sec:nonr_trans}
\subsection{Dispersion and Isofrequency Diagrams of the Unbounded Modulated Slab Medium}

In order to gain deeper insight into the wave propagation phenomenology within the space-time modulated slab medium, we next study the dispersion and isofrequency diagrams of the corresponding unbounded medium. Both are generally computed using~\eqref{eqa:det_gen} with~\eqref{eqa:K_matrix} and~\eqref{eqa:eps_sin_exp}.

\begin{figure}[h!]
\begin{center}
\includegraphics[width=1.06\columnwidth]{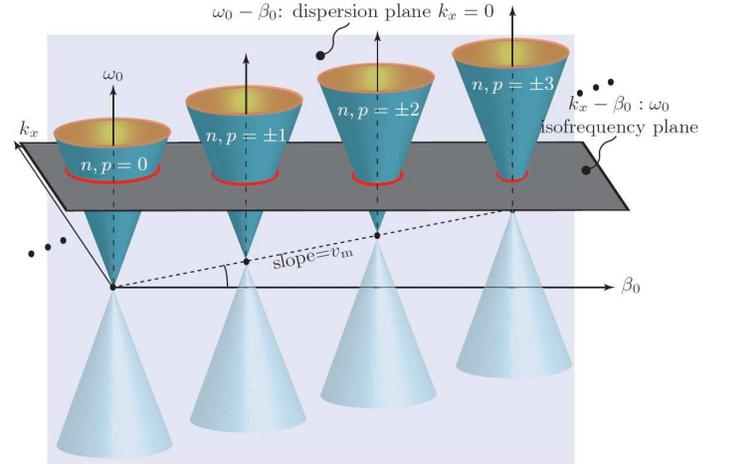}
\caption{Illustration of the three dimensional dispersion for the unbounded sinusoidally space-time modulated permittivity~\eqref{eqa:sin_perm}. A vertical cut at $k_x=0$ produces the dispersion diagrams, ($\omega_0,\beta_0$). A horizontal cut at the excitation frequency, $\omega_0$, produces isofrequency diagrams, ($\beta_0,k_x$). Note that the $\omega_0$, $\beta_0$ and $k_x$ axes are mutually orthogonal.}
\label{Fig:3D_dispersion}
\end{center}
\end{figure}

In the limiting case of a vanishingly small (but non zero) modulation depth, $\epsilon_\text{m}\rightarrow 0$, the aforementioned equations lead to the closed-form dispersion relation
\begin{equation}
k_x^2+(\beta_0^\pm \pm n \beta_\text{m})^2= \left(\frac{\omega_0+n\omega_\text{m}}{v_\text{b}} \right)^2.
\label{eqa:KKKK}
\end{equation}
Using~\eqref{eqa:vm_omegam_betam} and~\eqref{eqa:gamma}, this relation may be more conveniently rewritten as
\begin{equation}
\left(\frac{k_x}{\beta_\text{m}}\right)^2+\left(\frac{\beta_0^\pm }{\beta_\text{m}} \pm  n \right)^2
= \gamma^2 \left(\frac{\omega_0}{\omega_\text{m}} +n \right)^2,
\label{eqa:KKKK2}
\end{equation}
\noindent which represents an infinite periodic set of double cones with apexes at $k_x=0$ and $\beta_0=\pm n \beta_\text{m}$ and slope $v_\text{m}$, as illustrated in Fig.~\ref{Fig:3D_dispersion}. A vertical cross section of this 3D diagram at $k_x=0$ produces an infinite periodic set of straight lines in the $\beta_0/\beta_\text{m}-\omega_0/\omega_\text{m}$ plane, and a horizontal cut produces an infinite periodic set of circles centered at $(\beta_0^\pm/\beta_\text{m},k_x/\beta_\text{m})=(\mp n,0)$ with radius $\gamma \left(\omega_0/\omega_\text{m} +n \right)$ in the $\beta_0/\beta_\text{m}-k_x/\beta_\text{m}$ plane, as depicted in Fig.~\ref{Fig:3D_dispersion}.

\begin{figure}
\begin{center}
\includegraphics[width=0.95\columnwidth]{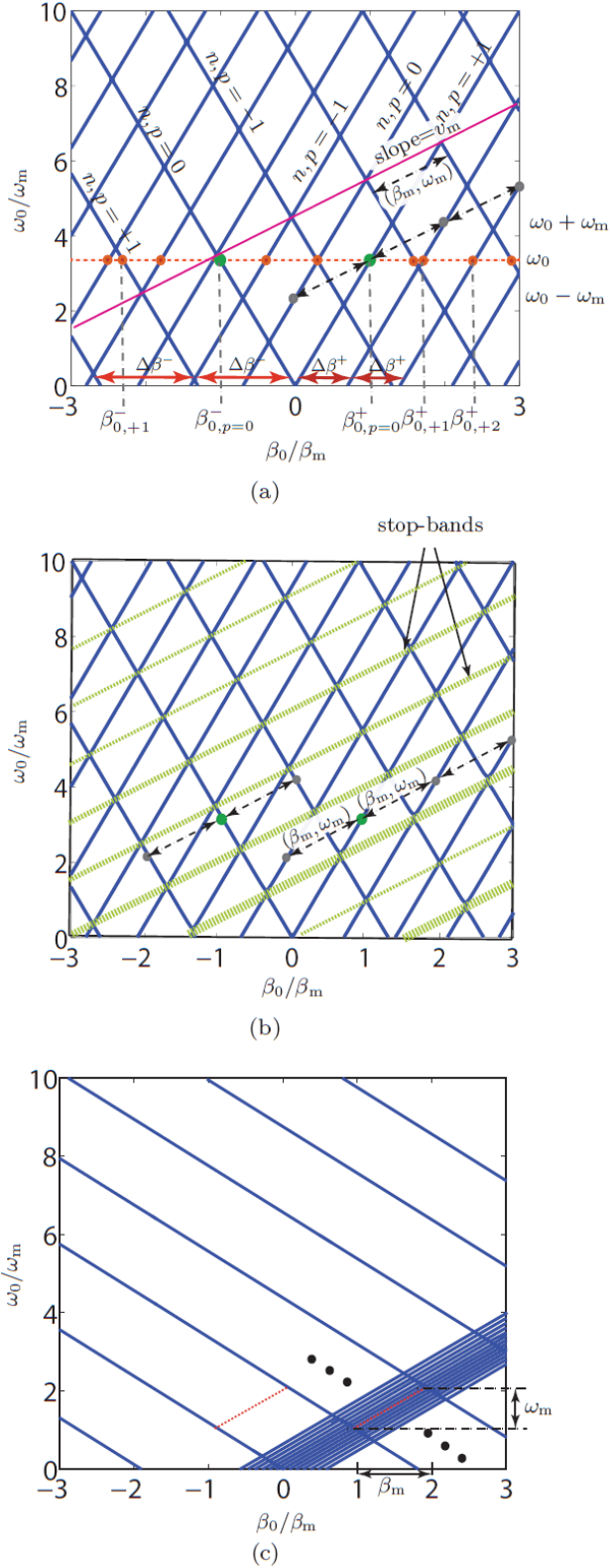}
\caption{Normal-incidence ($k_x=0$) dispersion diagrams for the sinusoidally space-time modulated (unbounded) slab medium with the permittivity~\eqref{eqa:sin_perm}, computed using~\eqref{eqa:det_gen} with~\eqref{eqa:K_matrix} and~\eqref{eqa:eps_sin_exp}.
(a)~Space-time modulated medium with vanishingly small
modulation depth, i.e. $\epsilon_\text{m}\rightarrow 0$ and for \mbox{$\gamma=0.3$ [Eq.~\eqref{eqa:KKKK2}]}.
(b)~Same as (a) except for the greater modulation depth $\epsilon_\text{m}=0.22\epsilon_\text{r}$. (c)~Same as (b) except for the subsonic ($\gamma<\gamma_\text{s,min}$) quasi-sonic ($\gamma\approx\gamma_\text{s,min}$) space-time modulation ratio $\gamma=0.85$ ($\gamma_\text{s,min}=0.905$).}
\label{Fig:dispersion_curves}
\end{center}
\end{figure}

We shall now consider the dispersion diagrams plotted in  Fig.~\ref{Fig:dispersion_curves} for the case of normal incidence, and therefore normal propagation everywhere, i.e. $k_x=0$, for different sets of parameters. Figure~\ref{Fig:dispersion_curves}(a) plots the dispersion diagram for a vanishingly small modulation depth, i.e. $\epsilon_\text{m}\rightarrow 0$, and for $\gamma=0.3$. In such a case, Eq.~\eqref{eqa:KKKK2} reduces to the very simple dispersion relation
\begin{equation}
\frac{\beta_0^\pm }{\beta_\text{m}}
= \gamma \frac{\omega_0}{\omega_\text{m}} +n \left( \gamma \mp  1\right).
\label{eqa:KKKK3}
\end{equation}
This diagram consists of the infinite periodic set of $\beta_0/\beta_\text{m}-\omega_0/\omega_\text{m}$ straight curves, labeled by $n$. To any frequency, $\omega_0$, corresponds an infinite number of modes, labeled by $p$, each of which consisting in the infinite number of forward and backward space-time harmonics $(\beta_{0p}^\pm+n\beta_\text{m},\omega_0+n\omega_\text{m}$) located on the corresponding oblique curve with slope $v_\text{m}$. Note that, as pointed out in Sec.~\ref{sec:gen_sol}, any (oblique) space-time harmonic point may be seen as corresponding to a different mode, excited at a different frequency, or, equivalently, that any mode, excited at $\omega_0$, may be seen as corresponding to an oblique space-time harmonic of another mode, with different excitation frequency. Given the vanishingly small periodic perturbation ($\epsilon_\text{m}\rightarrow 0$), the medium is here quasi homogeneous, with most of the energy residing in the $n=0$ forward and backward space-time harmonics, which would in fact represent the only remaining curves for exactly $\epsilon_\text{m}=0$ (homogeneous non-periodic medium).

We note in Fig.~\ref{Fig:dispersion_curves}(a) that when the velocity ratio is non-zero (here $\gamma=0.3$), the distances between the forward and backward space-time harmonics, $\Delta\beta^\pm=\beta_{n+1}^\pm-\beta_n^\pm$, are different. Specifically, as $\gamma$ increases, $\Delta\beta^+$ decreases and $\Delta\beta^-$ increases. This may be explained as follows, considering the horizontal cut $\omega_0=0$, where $\Delta\beta$ represents the spatial-frequency period or Brillouin zone edge. In the static case, $v_\text{m}=0$ (not shown in Fig.~\ref{Fig:dispersion_curves}), we have $\Delta\beta^\pm=\Delta\beta=2\pi/p_\text{stat}$, where $p_\text{stat}$ is the static period seen by both the forward and backward waves. As $v_\text{m}>0$ (all of Figs.~\ref{Fig:dispersion_curves}), the forward and backward waves see the velocities, $v^\pm=v_\text{b}\mp v_\text{m}$, respectively, \emph{relative to the modulating wave}, with limits $v^+(v_\text{m}=v_\text{b})=0$ and $v^-(v_\text{m}=v_\text{b})=2v_\text{b}$. The corresponding relative periods, satisfying the conditions $p_\text{mov}^\pm(v_\text{m}=0)=p_\text{stat}$, $p_\text{mov}^+(v_\text{m}=v_\text{b})=\infty$ (synchronization with modulation and hence no period seen, i.e. infinite period) and $p_\text{mov}^-(v_\text{m}=v_\text{b})=p_\text{stat}/2$ (due to opposite propagation at same velocity as modulation), are found as $p_\text{mov}^\pm=p_\text{stat}v_\text{b}/(v_\text{b}\mp v_\text{m})$. Thus, $\Delta\beta_\text{mov}^\pm=2\pi/p_\text{mov}^\pm=(2\pi/p_\text{stat})(v_\text{b}\mp v_\text{m})/v_\text{b}$ or $\Delta\beta_\text{mov}^\pm/\beta_m=1\mp\gamma$, indicating that distances between the forward and between the backward space-time harmonics decrease and increase, respectively, tending to the limits $\Delta\beta_\text{mov}^+(\gamma=1)/\beta_m=0$ and $\Delta\beta_\text{mov}^-(\gamma=1)/\beta_m=2$. This result, deduced from a physical argument, is in agreement with the mathematical result of~\eqref{eqa:KKKK3}, evaluating $\beta_{0,n+1}^\pm/\beta_\text{m} - \beta_{0,n}^\pm/\beta_\text{m}$ at $\omega_0=0$.

Figure~\ref{Fig:dispersion_curves}(b) plots the dispersion diagram for the greater modulation depth $\epsilon_\text{m}=0.22\epsilon_\text{r}$. In this case, the periodic perturbation is much more pronounced. Therefore, a substantial number of space-time harmonics contribute to the fields and the interferences between these harmonics are sufficiently strong to open up stop-bands. These stop-bands are naturally oblique, again with slope $v_\text{m}$, as they have to occur at the space-time synchronization points, i.e. at the intersections of the space-time harmonics which lie on oblique lines according to Fig.~\ref{Fig:dispersion_curves}(a). 

Figure~\ref{Fig:dispersion_curves}(c) plots the dispersion diagram for the greater space-time modulation ratio $\gamma=0.85$, which is subsonic ($\gamma<1$) and quasi-sonic ($\gamma\approx\gamma_\text{s,min}$) given $\gamma_\text{s,min}=0.905$. What is observed corresponds to the expectation from the above explanation and related formula $\Delta\beta_\text{mov}^\pm/\beta_m=1\mp\gamma$. The forward space-time harmonics get closer to each other (they would in fact completely fill up the diagram in the limit $\gamma\rightarrow 1$) and eventually collapse into a single curve at $\gamma=1$ since $\Delta\beta_\text{mov}^+(\gamma=1)/\beta_m=0$, producing the shock wave mentioned in Sec.~\ref{sec:gen_sol}. On the other hand, the backward space-time harmonics tend to be separated by the distance $\Delta\beta_\text{mov}^-(\gamma=1)/\beta_m=2$. In this quasi-sonic regime, the closest space-time harmonics strongly couple to each other at a given frequency $\omega_0$ because they possess very close phase velocities and are hence essentially phase-matched to each other. Therefore, this is regime of particular interest, as will be seen in the application of Sec.~\ref{sec:isolator}. As mentioned in Sec.~\ref{sec:gen_sol}, the analytical results presented in this paper are restricted to the subsonic regime, but the quasi-sonic condition $\gamma\approx\gamma_\text{s,min}$ allows one to reap the essential benefits of the physics occurring in the sonic regime, as will be seen in Sec.~\ref{sec:isolator}.

\begin{figure}
\begin{center}
\includegraphics[width=1.05\columnwidth]{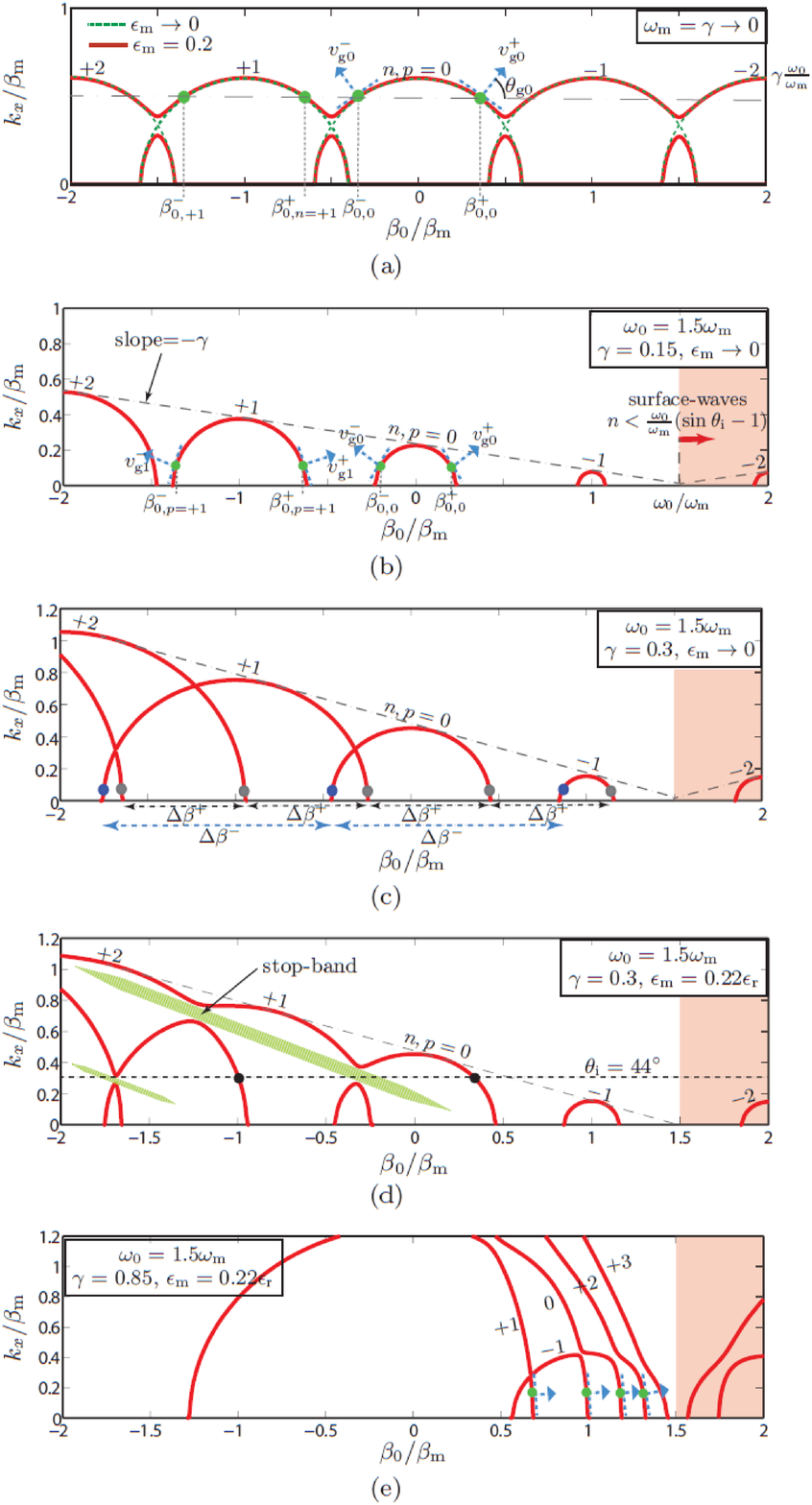}
\caption{Isofrequency diagrams for the unbounded sinusoidally space-time modulated medium~\eqref{eqa:sin_perm}, computed using~\eqref{eqa:det_gen} with~\eqref{eqa:K_matrix} and~\eqref{eqa:eps_sin_exp}.
(a)~Purely space-modulated medium, i.e. $\omega_\text{m}=\gamma\rightarrow0$ (but finite $\gamma \omega_\text{0}/\omega_\text{m}=0.6$).
(b)~Space-time modulated medium with vanishingly small modulation depth, i.e. $\epsilon_\text{m}\rightarrow 0$, and for $\omega_0=1.5\omega_\text{m}$ and $\gamma=0.15$.
(c)~Same as (b) except for the greater space-time modulation ratio $\gamma=0.3$.
(d)~Same as (c) except for the space-time modulation depth $\epsilon_\text{m}=0.22\epsilon_\text{r}$.
(e)~Same as (d) but in the quasi-sonic regime with $\gamma=0.85$ ($\gamma_\text{s,min}=0.905$).}
\label{Fig:2D_dispesion_curves}
\end{center}
\end{figure}

We shall now consider the isofrequency diagrams plotted in  Fig.~\ref{Fig:2D_dispesion_curves} for different sets of parameters. These diagrams may be easiest understood using the 3D perspective  in Fig.~\ref{Fig:3D_dispersion}. Figure~\ref{Fig:2D_dispesion_curves}(a) plots the isofrequency curves for a purely space modulated medium, where $\omega_\text{m}=\gamma\rightarrow 0$, with vanishingly small modulation depth, i.e. $\epsilon_\text{m}\rightarrow 0$, and $\epsilon_\text{m}=0.2$. In the former case ($\epsilon_\text{m}\rightarrow 0$), the curves are the circles given by~\eqref{eqa:KKKK2}, corresponding to the infinite number of space harmonics $n$ and reducing to the center circle, $k_x^2+\beta_0^2=k^2-k_0^2/\epsilon_\text{r}$, in the trivial limiting case of a perfectly homogenous medium ($\epsilon_\text{m}=0$). In the latter case ($\epsilon_\text{m}=0.2$), Eq.~\eqref{eqa:KKKK2} is not valid any more and one must resort to the general relation~\eqref{eqa:det_gen}. Here, spatial ($k_x-\beta$) stop-bands open up at the intersection points for $\epsilon_\text{m}=0$, due to space harmonic coupling. Note that in such a purely spatially modulated medium, the space harmonics are simply related by $\beta_{0,n}^\pm=\beta_{0,0}^\pm +n\beta_\text{m}$, where, for a given $k_x$, all the spatial harmonics propagate, attenuate (near stop-bands edges, when $\epsilon_\text{m} > 0$) or get cut off.


Figure~\ref{Fig:2D_dispesion_curves}(b) plots the isofrequency diagram for a space-time modulated medium, where $\beta_\text{m},\omega_\text{m}> 0$, still with vanishingly small modulation depth, i.e. $\epsilon_\text{m} \rightarrow 0$, and $\omega_0=1.5\omega_\text{m}$, $\gamma=0.15$. The isofrequency circles have now different radii, corresponding to $\gamma \left(\omega_0/\omega_\text{m} +n \right)$[Eq.~\eqref{eqa:KKKK2}], with envelope slope of $-\gamma$. This is because, in contrast to the purely spatial medium in Fig.~\ref{Fig:2D_dispesion_curves}(a), the space-time medium supports, for fixed incidence angle (i.e. fixed $k_x$), an infinite number of modes (labeled by $p$), each of them composed of an infinite number of forward and backward space-time harmonics $(\beta_{0p}^\pm+n\beta_\text{m},\omega_0+n\omega_\text{m}$), as previously explained. The medium operates as a spatial ($\beta$) high-pass filter, some modes above the cutoff propagating and those below the cut off, as shown in Fig.~\ref{Fig:S-T_slab_2D} and related explanation.

Figure~\ref{Fig:2D_dispesion_curves}(c) shows the same diagram as Fig.~\ref{Fig:2D_dispesion_curves}(b), for the larger space-time modulation ratio, $\gamma=0.3$. As $\gamma$ increases, the isofrequency circles in the forward region, $\beta_0>0$, get smaller and appear farther apart. In contrast, those in the backward region, $\beta_0<0$, get bigger, and eventually intersect. This behavior can be intuitively understood from the 3D dispersion curves in Fig.~\ref{Fig:3D_dispersion}.

Figure~\ref{Fig:2D_dispesion_curves}(d) shows isofrequency curves for the stronger modulation depth $\epsilon_\text{m}=0.22\epsilon_\text{r}$. As the modulation depth is increased, forward and backward waves at the intersections of the isofrequency circles couple more strongly and a visible bandstop appears in the isofrequency diagram. In the limit of vanishingly small modulation depth, these bandstops become vanishingly narrow. Depending on the incidence frequency and angle, some modes propagate and some are cut off. For instance, for $\theta_\text{i}=44^\circ$, mode $p=-1$ is evanescent, while modes $p=0$ and $p=+2$ represent forward propagating waves.


Finally, Fig.~\ref{Fig:2D_dispesion_curves}(e) plots the isofrequency diagram in the quasi-sonic regime. The forward waves are synchronized and exhibit similar group and phase velocities, leading to strong interaction and coupling, while backward waves are more distant and therefore interact relatively weakly.

%

%
\subsection{Nonreciprocal Scattering from the Slab}
This section studies the nonreciprocity of the space-time modulated system in Fig.~\ref{Fig:S-T_slab}. The structure is analyzed with the analytical technique presented in Sec.~\ref{sec:sin_mod} and verified using full-wave finite difference frequency domain (FDTD) simulations.

\begin{figure}[h!]
\begin{center}
\includegraphics[width=0.95\columnwidth]{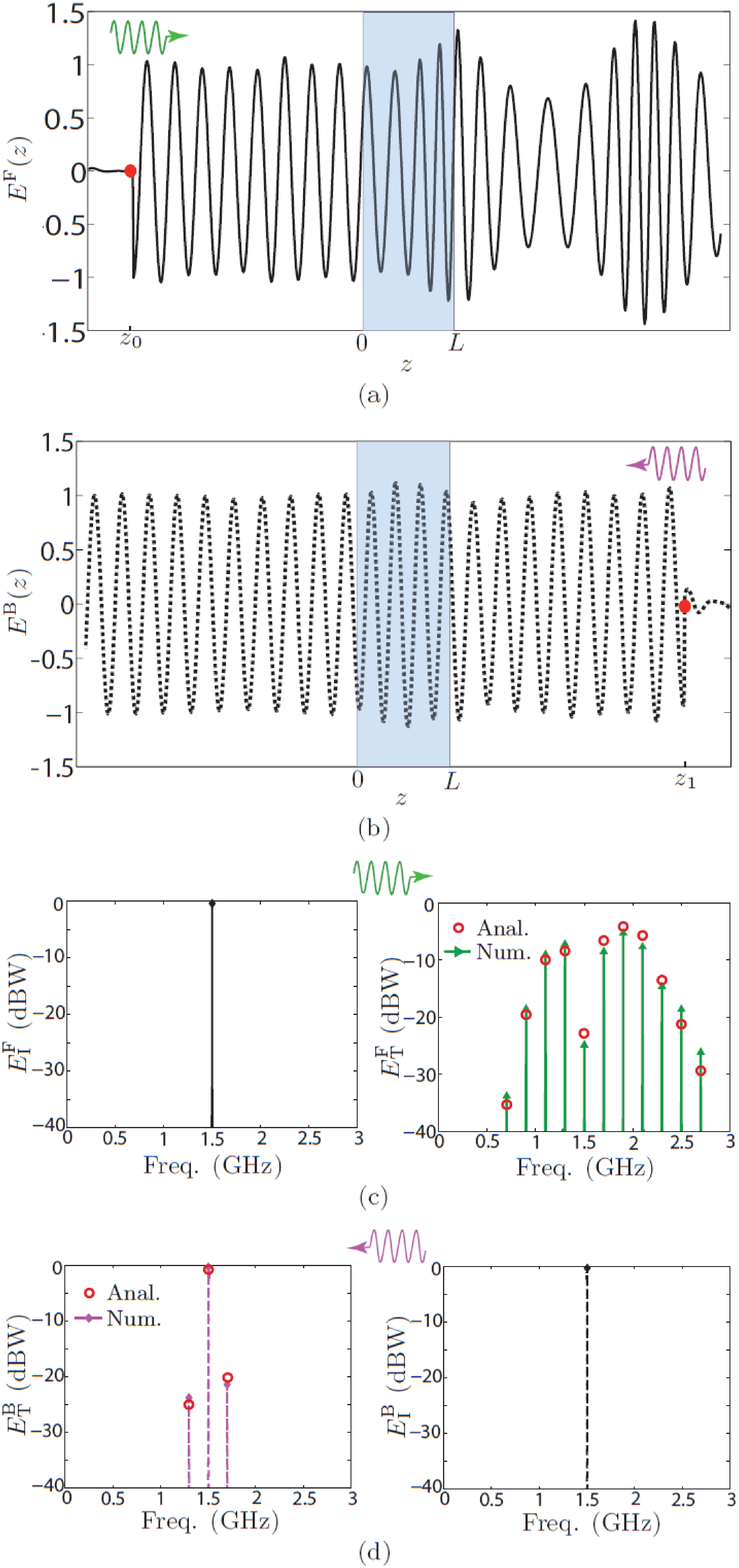}
\caption{Analytical [Eqs.~\eqref{eqa:E_TF} and~\eqref{eqa:E_TB}] and numerical (FDTD) results for the forward and backward problems in the quasi-sonic regime with parameters $\epsilon_\text{m}=0.3\epsilon_\text{r}$, $\omega_0=2\pi\times1.5$~GHz, $\omega_\text{m}=2\pi\times 0.2$~GHz, $L=3\lambda_0$ and $\gamma=0.85$. (a)~and (b) FDTD waveforms showing the electric field amplitude for the forward and backward problems, respectively. (c)~and (d) Temporal frequency spectrum of the transmitted field for the forward and backward problems, respectively.}
\label{Fig:Nonrec_trans}
\end{center}
\end{figure}

First consider the forward problem in Fig.~\ref{Fig:S-T_slab}. A wave is normally incident on the slab with sinusoidal permittivity~\eqref{eqa:sin_perm} and operated in the quasi-sonic regime with  velocity ratio $\gamma=0.85$ ($\gamma_\text{s,min}=0.867$). Figure~\ref{Fig:Nonrec_trans}(a) shows the FDTD response for the amplitude of the electric field. The wave strongly interacts with the medium as it passes through the slab and generates all the space-time harmonics. The corresponding temporal frequency spectrum for the transmitted wave is plotted in Fig.~\ref{Fig:Nonrec_trans}(c). The incident power at $\omega_0$ is effectively converted in the space-time harmonics $\omega_0\pm n\omega_\text{m}$, $n\ge1$, yielding a transmitted wave carrying weak power at the incident frequency $\omega_0$.


Next, consider the backward problem. Figure~\ref{Fig:Nonrec_trans}(b) shows the FDTD response for the amplitude of the electric field. The wave weakly interacts with the medium as it passes through the slab and remains almost unaltered. The corresponding temporal-frequency spectrum, plotted in Fig.~\ref{Fig:Nonrec_trans}(c), confirms this fact. The two weak harmonics at $\omega_\text{0}+\omega_\text{m}$ and $\omega_\text{0}-\omega_\text{m}$ are due to local impedance mismatch [Eq.~\eqref{eqa:sin_imped}] in the medium.

The space-time medium affects forward and backward waves differently, producing strong harmonics in the forward problem and almost no harmonics in the backward problem. This nonreciprocity is exploited in Sec.~\ref{sec:isolator} for realizing of a quasi-sonic isolator. It should be noted that both in the forward and backward problems the incident wave couples to an infinite number space-time harmonics. However, in the backward problem this coupling is extremely weak.

\FloatBarrier

\begin{figure}
\begin{center}
\includegraphics[width=1\columnwidth]{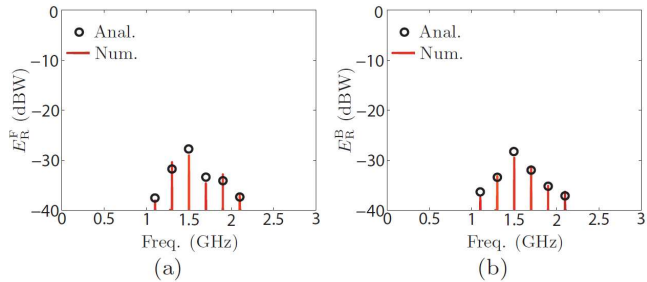}
\caption{Temporal frequency spectrum of the reflected field from a space-time modulated slab with the same parameters as in Fig.~\ref{Fig:Nonrec_trans}. (a)~Forward problem, $E_\text{R}^\text{F}$. (b)~Backward problem, $E_\text{R}^\text{B}$.}
\label{Fig:Nonrec_refl}
\end{center}
\end{figure}

Figures~\ref{Fig:Nonrec_refl}(a) and ~\ref{Fig:Nonrec_refl}(b) plot temporal frequency spectrum of the reflected wave, for the same parameters as in Fig.~\ref{Fig:Nonrec_trans}, for forward and backward excitations, respectively. In both cases the structure is well matched and reflects weakly. The reflection level is directly proportional to the modulation depth.


%

\subsection{Effect of the Velocity Ratio}

\begin{figure}[h!]
\begin{center}
\includegraphics[width=0.9\columnwidth]{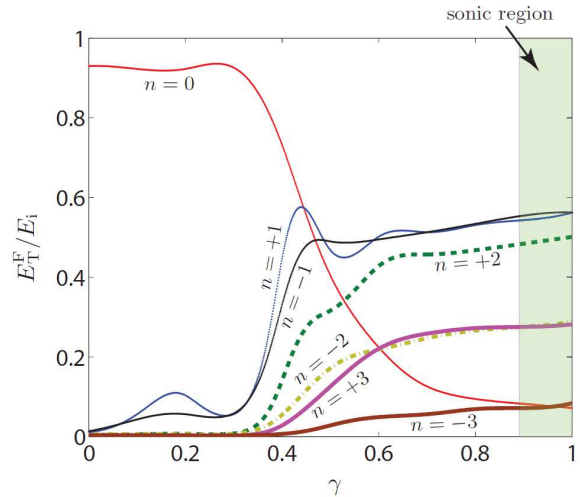}
\caption{FDTD (FFT) transmitted field versus velocity ratio ($\gamma$) showing the harmonics distribution for a space-time slab with parameters $\epsilon_\text{m}=0.22\epsilon_\text{r}$ ($\gamma_\text{s,min}=0.905$), $\omega_\text{m}=2\pi \times 0.2$~GHz, $\omega_\text{0}=2\pi \times 1.5$~GHz and $L=3.5\lambda_0$.}
\label{Fig:var_gamma}
\end{center}
\end{figure}

This section investigates the effect of the velocity ratio on the power conversion efficiency of the space-time modulated slab, i.e. the amount of power that is transmitted to desired space-time harmonics $\omega_0\pm n\omega_\text{m}$, $n\ge 1$. Figure~\ref{Fig:var_gamma} shows the distribution of the transmitted power in different harmonics versus the velocity ratio $\gamma$. The highlighted region represents the sonic regime. In the subsonic regime, where $\gamma~\rightarrow~0$, little energy is transferred to harmonics. As $\gamma$ approaches the sonic regime, power conversion efficiency is increased. In the quasi-sonic and sonic regimes most of the power is transferred to other harmonics with only a small amount of power remaining in the fundamental ($n=0$).

%
%

\begin{figure}
\begin{center}
\includegraphics[width=1.05\columnwidth]{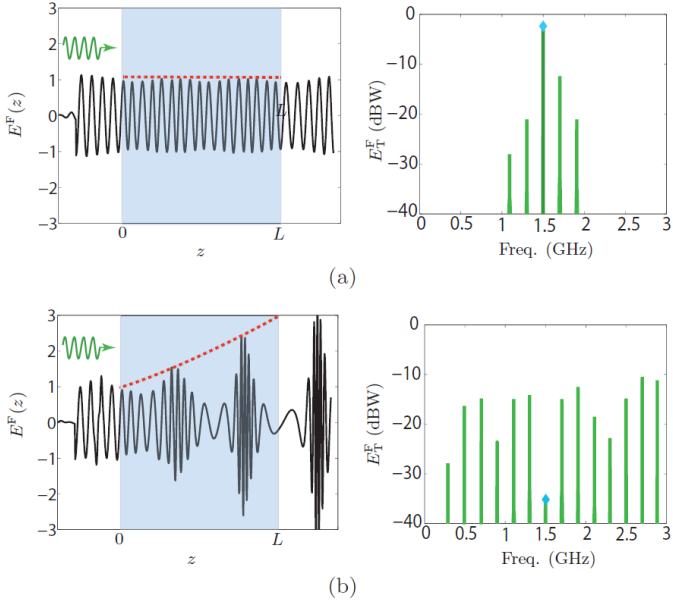}
\caption{FDTD comparison of the rate of power growth (forward problem) in subsonic and quasi-sonic regimes for (a)~The subsonic space-time velocity ratio $\gamma=0.3$ and (b)~The quasi-sonic space-time velocity ratio $\gamma=0.85$ ($\gamma_\text{s,min}=0.905$), where $\epsilon_\text{m}=0.22\epsilon_\text{r}$, $\omega_\text{m}=2\pi \times 0.2$~GHz, $\omega_\text{0}=2\pi \times 1.5$~GHz and $L=15\lambda_0$.}
\label{Fig:scatt_sonic}
\end{center}
\end{figure}

In the quasi-sonic and sonic regimes, the total wave power grows quasi-exponentially as the wave propagates along the space-time modulated section. Figure.~\ref{Fig:scatt_sonic} compares the wave amplitudes in subsonic and quasi-sonic regimes for $\gamma=0.3$ and $\gamma=0.85$, respectively, versus the position. In the subsonic regime the wave and the modulation are not synchronized and there is only weak coupling between the two. As a result the wave magnitude is almost flat. In contrast, in the quasi-sonic and sonic regimes the wave and modulation velocities are synchronized. Consequently, the two are strongly coupled and the wave amplitude grows quasi-exponentially along the space-time modulated slab. This observation is not at odds with power conservation, as energy is pumped into the system through space-time modulation. However, this exponential growth can not be efficiently used for wave amplification, since power is distributed among infinite space-time harmonics, as seen in the temporal frequency spectrum in Fig.~\ref{Fig:scatt_sonic}(b).

%
%

\subsection{Effect of the Length}

\begin{figure}
\begin{center}
\includegraphics[width=0.9\columnwidth]{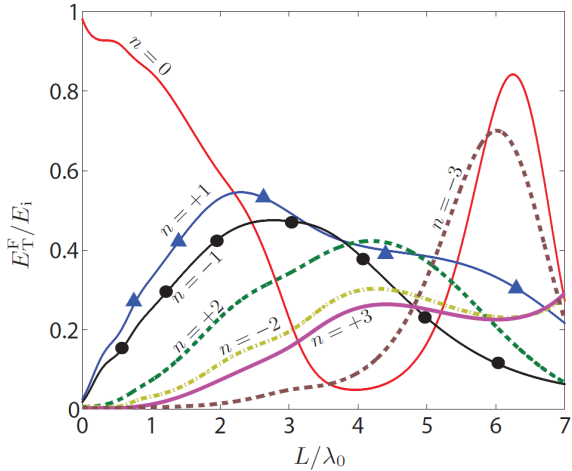}
\caption{Magnitude of different harmonics versus the length of the space-time slab (FDTD). The slab operates in the middle of sonic regime with parameters $\epsilon_\text{m}=0.22\epsilon_\text{r}$, $\omega_\text{m}=2\pi \times 0.2$~GHz, $\omega_\text{0}=2\pi \times 1.5$~GHz.
}
\label{Fig:var_L}
\end{center}
\end{figure}

Figure~\ref{Fig:var_L} plots the magnitude of first harmonics with respect to the length of the space-time modulated slab. The structure operates in the middle of the sonic regime, i.e. $\gamma=1$. As the length of the slab is increased, the input power is more efficiently coupled to the space-time harmonics $\omega_0 \pm n\omega_\text{m}$, $n\ge1$. The incident wave gradually couples its energy to these harmonics as it propagates along the slab. Therefore, a longer slab exhibits more efficient power conversion. However, this effect saturates at some point as the power in these higher order modes couples back and transfers its energy back to the fundamental mode. Therefore, power conversion efficiency shows a quasi-periodic behavior with respect to the length of the slab.


\section{QUASI-SONIC ISOLATOR}\label{sec:isolator}
\begin{figure}
\centering
\psfrag{A}[c][c][0.9]{$\omega$}
\psfrag{B}[c][c][0.9]{$\beta$}
\psfrag{c}[c][c][0.9]{$\beta_{0}$-$\beta_\text{m}$}
\psfrag{e}[c][c][0.9]{$\beta_{0}$+$\beta_\text{m}$}
\psfrag{r}[c][c][0.9]{$\beta_{0}$-2$\beta_\text{m}$}
\psfrag{s}[c][c][0.9]{$\beta_{0}$+2$\beta_\text{m}$}
\psfrag{d}[c][c][0.9]{$\beta_{0}$}
\psfrag{g}[c][c][0.9]{$\beta_\text{m}$}
\psfrag{h}[l][c][0.9]{$\omega_\text{m}$}
\psfrag{i}[r][c][0.9]{$\omega_0$}
\psfrag{j}[r][c][0.9]{$\omega_0$+$\omega_\text{m}$}
\psfrag{k}[r][c][0.9]{$\omega_0$+2$\omega_\text{m}$}
\psfrag{l}[r][c][0.9]{$\omega_0$-$\omega_\text{m}$}
\psfrag{m}[r][c][0.9]{$\omega_0$-2$\omega_\text{m}$}
\psfrag{q}[l][l][0.9]{light cone}
\psfrag{n}[c][c][0.9]{-$\beta_{0}$}
\psfrag{o}[c][c][0.9]{-$\beta_{0}$+$\beta_\text{m}$}
\psfrag{p}[c][c][0.9]{-$\beta_{0}$-$\beta_\text{m}$}
\psfrag{K}[c][c][0.9]{\shortstack{Dispersion engineered\\first higher-order\\-leaky-mode}}
\psfrag{Q}[c][c][0.9]{\shortstack{no mode\\for transition}}
\centering{\includegraphics[width=\columnwidth] {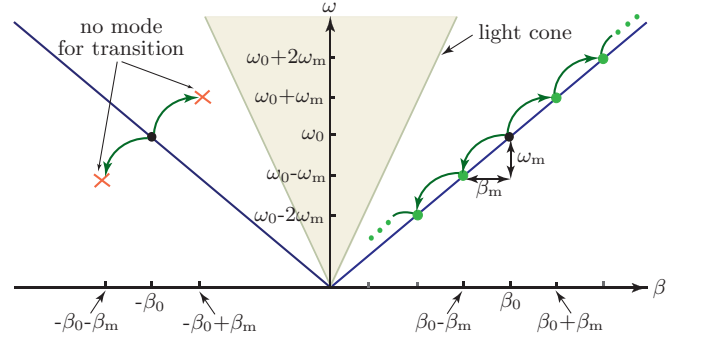}}
\caption{A space-time modulated slab operated in the quasi-sonic or sonic regime. In the forward direction the incident energy is transferred in cascade to space-time harmonics. In the backward direction the wave passes through with little interaction.
}
\label{Fig:disp_diag}
\end{figure}

\begin{figure}
\begin{center}
\includegraphics[width=\columnwidth]{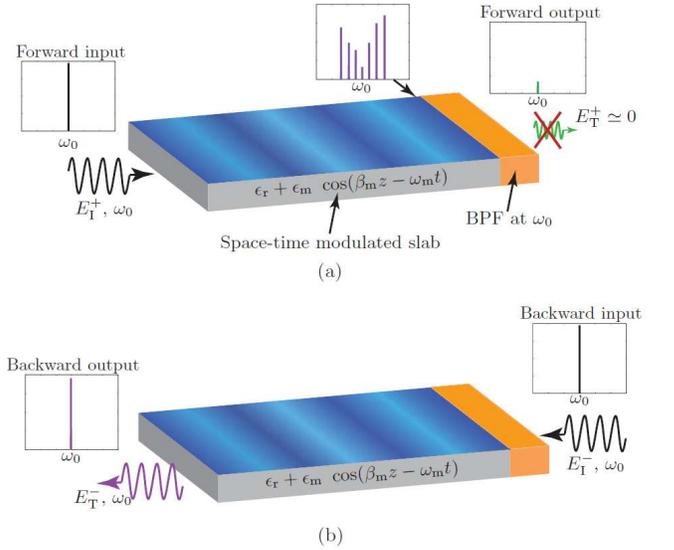}
\caption{Quasi-sonic isolator. The quasi-sonic or sonic space-time modulated slab in Fig.~\ref{Fig:disp_diag} is connected to a bandpass filter. (a)~In the forward direction, the incident power is converted to higher order harmonics and eliminated by the filter (BPF). (b)~In the backward direction, the wave passes through the system.
}
\label{Fig:iso_generic}
\end{center}
\end{figure}

This section exploits the strong nonreciprocity of the quasi-sonic and sonic regimes for the realization of an electromagnetic isolator. The principle of operation of the proposed isolator is illustrated in Figs.~\ref{Fig:disp_diag} and \ref{Fig:iso_generic}. A space-time slab is operated in the quasi-sonic or sonic regime. The length and modulation ratio are adjusted such that in the forward direction the incident power is efficiently converted to higher order space-time harmonics $\omega_0\pm\omega_\text{m}$, $n\ge1$,  little energy is transmitted at the fundamental frequency $\omega_0$. In contrast, in the backward direction the space-time slab interacts weakly with the incident wave and therefore the incident wave passes through almost unaltered. If the transmitted wave is passed through a bandpass filter with bandpass frequency $\omega_0$, as in Fig.~\ref{Fig:iso_generic}, in the forward direction most of the power is in the stopband and is therefore dissipated or reflected by the filter. However, in the backward direction most of the power is at fundamental frequency $\omega_0$ and passes through. Thus, the structure operates as an isolator.

%

%

%

%

\begin{figure}
\begin{center}
\includegraphics[width=\columnwidth]{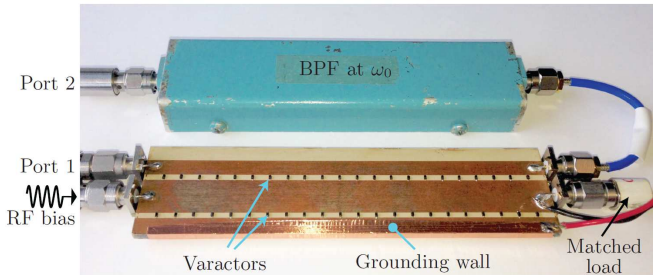}
\caption{Photograph of the fabricated isolator, employing SMV1247 varactors manufactured by Skyworks Solutions with capacitance ratio $C_\text{max}/C_\text{min} = 10$. The specifications of the structure are $L=11.7$~cm, RT6010 substrate with permittivity~$10.2$, thickness~$h = 100$~mil and $\tan \delta = 0.0023$.}
\label{Fig:photo}
\end{center}
\end{figure}

We realized the space-time modulated slab with permittivity~\eqref{eqa:sin_perm} using a microstrip transmission line loaded with an an array of sub-wavelengthly spaced varactors. The fabricated prototype is shown in Fig.~\ref{Fig:photo}. The varactors are reversed biased by a DC voltage and are spatio-temporally modulated by an RF bias, realizing the space-time varying capacitance $C(z,t)= C_\text{av}+C_\text{m} ~\cos(\beta_\text{m}z-\omega_\text{m}t)$. This circuit emulates a medium with effective permittivity $\epsilon(z,t)= \epsilon_\text{av}+\epsilon_\text{m} ~\cos(\beta_\text{m}z-\omega_\text{m}t)$, with $\epsilon_\text{av}=\epsilon_\text{ef}+\epsilon_\text{av,var}$, where $\epsilon_\text{ef}$ is the effective permittivity of the microstrip line and $\epsilon_\text{av,var}$ is the average permittivity introduced by the varactors. The modulation depth is controlled through the amplitude of the RF bias.

\begin{figure}
\begin{center}
\includegraphics[width=0.95\columnwidth]{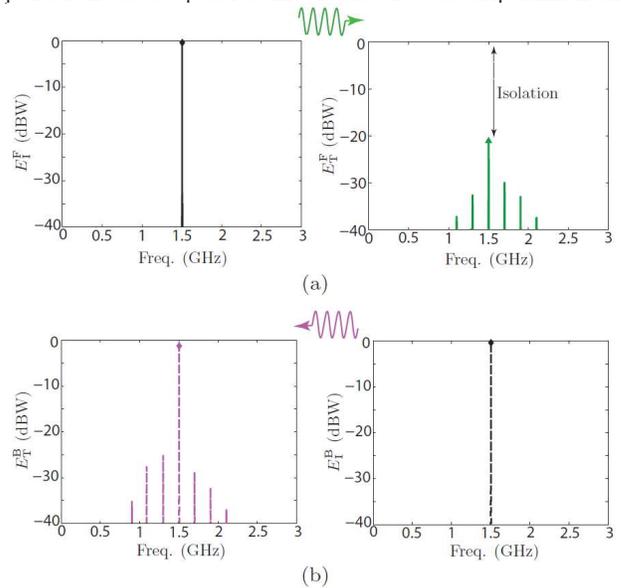}
\caption{Experimental results for the isolator in Figs.~\ref{Fig:iso_generic} and~\ref{Fig:photo} with $\epsilon_\text{m}=0.22\epsilon_\text{r}$, $\omega_0=2\pi\times1.5$ GHz, $\omega_\text{m}=2\pi\times 0.2$ GHz, $L=3.5\lambda_0$ and $\gamma\rightarrow1$. (a)~Forward problem. (b)~Backward problem.}
\label{Fig:iso_res}
\end{center}
\end{figure}

\begin{figure}
\begin{center}
\subfigure[]{\label{Fig:Nonrec_refl_exp_a}
\psfrag{C}[l][c][0.7]{Anal.}
\psfrag{D}[l][c][0.7]{Num.}
\psfrag{A}[c][c][0.7]{Freq. (GHz)}
\psfrag{b}[c][c][0.7][90]{$E_\text{R}^\text{F}$ (dBW)}
\psfrag{u}[c][c][0.85]{Forward case reflection}
\includegraphics[width=0.45\columnwidth]{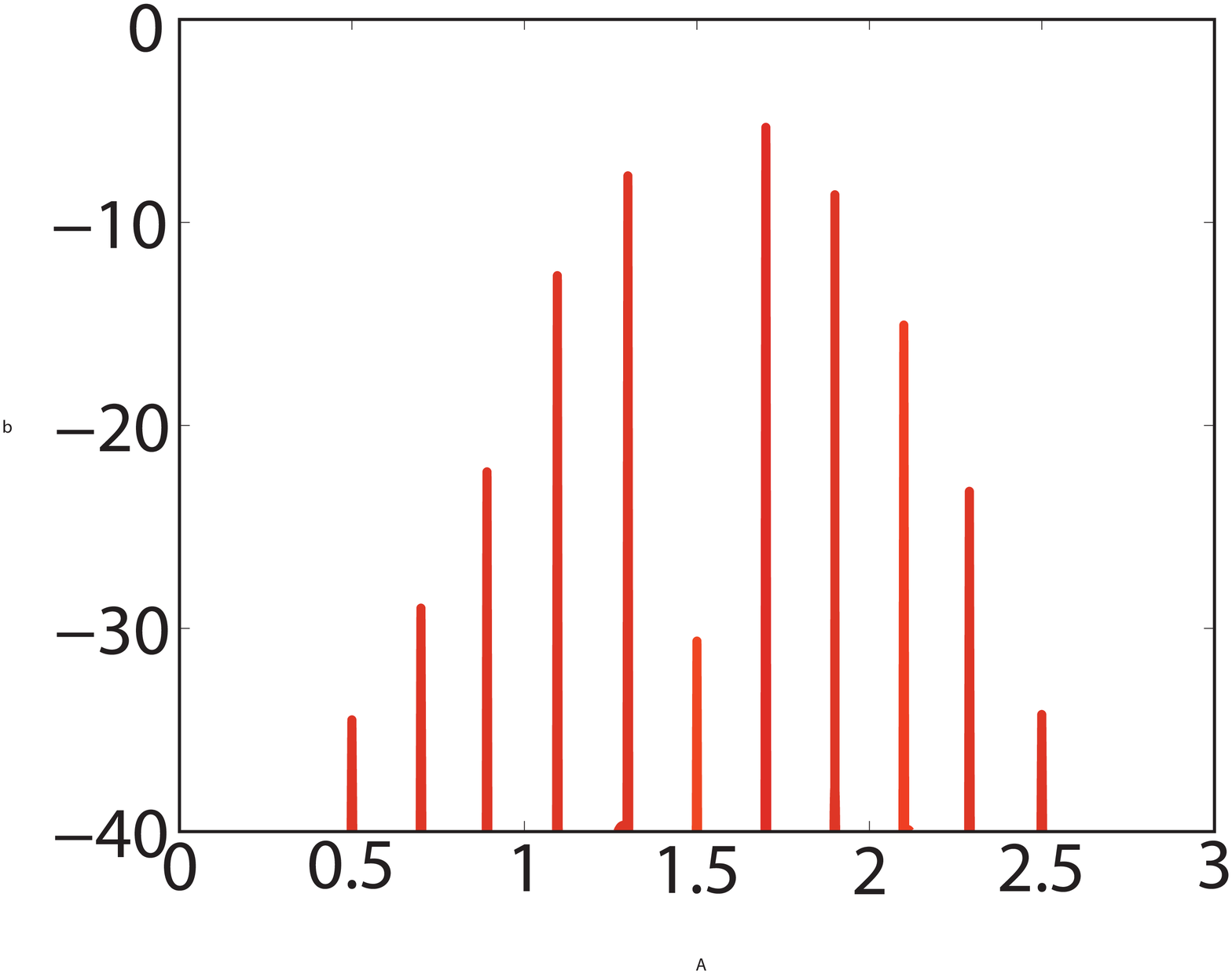}}
\subfigure[]{\label{Fig:Nonrec_refl_exp_b}
\psfrag{u}[c][c][0.85]{Backward case reflection}
\psfrag{C}[l][c][0.7]{Anal.}
\psfrag{D}[l][c][0.7]{Num.}
\psfrag{A}[c][c][0.7]{Freq. (GHz)}
\psfrag{b}[c][c][0.7][90]{$E_\text{R}^\text{B}$ (dBW)}
\includegraphics[width=0.45\columnwidth]{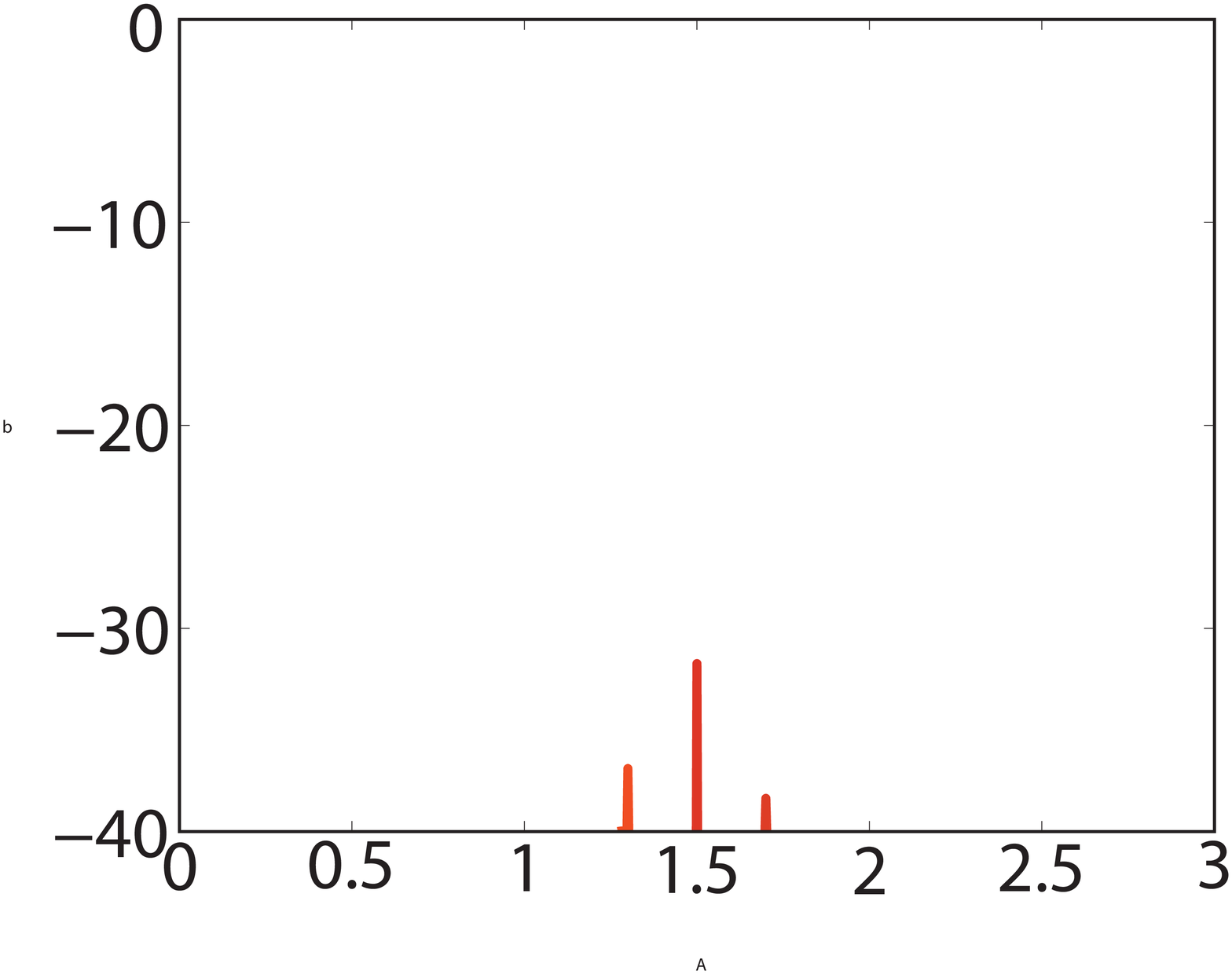}}
\caption{Measured reflections from the isolator slab in Figs.~\ref{Fig:iso_generic} and~\ref{Fig:photo}, with the same parameters as in Fig.~\ref{Fig:iso_res}. (a)~Forward problem. (b)~Backward problem.}
\label{Fig:Nonrec_refl_exp}
\end{center}
\end{figure}

Figures~\ref{Fig:iso_res}(a) and~\ref{Fig:Nonrec_refl_exp} show the measurement results. The space-time varying microstip circuit was connected to a bandpass filter, and forward and backward transmission and reflection coefficients were measured. In the forward direction, corresponding to Fig.~\ref{Fig:iso_res}(a), the transmission level is less than $-20$~dB at the fundamental harmonic, and less than $-30$~dB in other spacetime harmonics. Figure~\ref{Fig:Nonrec_refl_exp}(a) shows that the power injected into the higher order harmonics $\omega_0 \pm n\omega_\text{m}$ is reflected by the bandpass filter. In the backward direction, shown in Fig.~\ref{Fig:iso_res}(b), the incident wave is almost fully transmitted at the fundamental frequency, with less than $-30$~dB reflection. Thus, the structure realizes an isolator with more than $20$~dB isolation.

\section{CONCLUSION}\label{sec:conclusion}
We studied the scattering of obliquely incident electromagnetic waves from periodically space-time modulated slabs. It is shown that such a structure operates as a nonreciprocal harmonic generator. We showed that the structure operates as a high pass filter in oblique incidence, where low frequency harmonics are filtered out in the form of surface waves, while high frequency harmonics are scattered as space waves. In the quasi-sonic regime, where the velocity of space-time modulation is close to the velocity of the electromagnetic waves in the background medium, the incident wave is strongly coupled to space-time harmonics in the forward direction while in the backward direction it exhibits low coupling to other harmonics. This nonreciprocity has been leveraged for realization of an electromagnetic isolator in the quasi-sonic regime. The space-time varying medium was realized at microwave frequencies using a microstrip line loaded with sub-wavelengthly spaced varactors, and its operation as a quasi-sonic isolator was experimentally demonstrated.


\bibliographystyle{IEEEtran}
\bibliography{Taravati_Space_Time_Reference}

\pagebreak

\onecolumngrid

\newpage

  \renewcommand{\theequation}{S\arabic{equation}}
  \renewcommand{\thesubsection}{\arabic{subsection}}
    \renewcommand{\thesubsubsection}{3.\arabic{subsubsection}}
  \setcounter{equation}{0}  
  \setcounter{subsection}{0}  

\headsep = 40pt

\section*{Supplemental~Material for\\``Electromagnetic Scattering from a Periodically Space-Time Modulated Slab \\ and Application to a Quasi-Sonic Isolator''}

\textbf{\subsection{Sonic-Regime Interval}}\label{sec:A-sonic}
The sonic-regime interval was shown to correspond to a singularity and mathematically determined for the particular case of a one-dimensional sinusoidally modulated medium in~\cite{Oliner_Hessel_1959}. Following the same approach, we derive here the sonic-regime interval for the case of a two-dimensional general periodic modulated medium.

We first expand the wave equation inside the modulated slab, given by~\eqref{eqa:wave_eq}, as
\begin{equation}\label{eqa:A-wave_eq}
c^2 \frac{{{\partial ^2}\mathbf{E}_\text{M}}}{{\partial {x^2}}}+c^2 \frac{{{\partial ^2}\mathbf{E}_\text{M}}}{{\partial {z^2}}} = \frac{{{\partial ^2} \left[\epsilon (z,t)\mathbf{E}_\text{M} \right]}}{{\partial {t^2}}}
 = \mathbf{E}_\text{M} \frac{{{\partial ^2} \epsilon (z,t) }}{{\partial {t^2}}} +\epsilon (z,t) \frac{{{\partial ^2} \mathbf{E}_\text{M} }}{{\partial {t^2}}} +2\frac{{{\partial } \epsilon (z,t) }}{{\partial {t}}} \frac{{{\partial \mathbf{E}_\text{M} } }}{{\partial {t}}},
\end{equation}
\noindent where $\mathbf{E}_\text{M}=\mathbf{E}_\text{M}(x,z,t)$. Then, we apply the moving medium coordinate transformation
\begin{equation}\label{eqa:A-cord_transf}
  x'=x, \qquad    u=-\beta_\text{m}z+\omega_\text{m}t, \qquad  t'=t.
\end{equation}
Next we express the partial derivatives in~\eqref{eqa:A-wave_eq} in terms of the new variable in~\eqref{eqa:A-cord_transf}, i.e.
\begin{subequations}\label{eqa:A-wave_transf}
\begin{equation}
\frac{\partial }{\partial x}=\frac{\partial {x'}}{\partial x} \frac{\partial }{\partial x'}+\frac{\partial u}{\partial x} \frac{\partial }{\partial u}+\frac{\partial t'}{\partial x} \frac{\partial }{\partial t'}=  \frac{\partial }{\partial x'},
\end{equation}
\begin{equation}
\frac{\partial }{\partial z}=\frac{\partial {x'}}{\partial z} \frac{\partial }{\partial x'}+\frac{\partial {u}}{\partial z} \frac{\partial }{\partial u}+\frac{\partial t'}{\partial z} \frac{\partial }{\partial t'}= -\beta_\text{m} \frac{\partial }{\partial u},
\end{equation}
\begin{equation}
\frac{\partial^{2} }{\partial z^{2}}= \frac{\partial }{\partial z}\left(\frac{\partial}{\partial z}\right) =  \beta_\text{m}^2 \frac{\partial^2 }{\partial u^2},
\end{equation}
\begin{equation}
\frac{\partial }{\partial t}= \frac{\partial x'}{\partial t} \frac{\partial }{\partial x'}  +\frac{\partial u}{\partial t} \frac{\partial }{\partial u}+\frac{\partial t'}{\partial t} \frac{\partial }{\partial t'} = \frac{\partial }{\partial t'}+\omega_\text{m} \frac{\partial }{\partial u},
\end{equation}
\begin{equation}
\frac{\partial^{2} }{\partial t^{2}}= \left(\frac{\partial {t'}}{\partial t}\right)^2 \frac{\partial^2 }{\partial t'^2}+\left(\frac{\partial u}{\partial t}\right)^2 \frac{\partial^2 }{\partial u^2}+2 \left(\frac{\partial t'}{\partial t}\right) \frac{\partial u}{\partial t} \frac{\partial^2 }{\partial u \partial t'} = \frac{\partial^2 }{\partial t'^2}+\omega_\text{m}^2 \frac{\partial^2 }{\partial u^2}+2 \omega_\text{m} \frac{\partial^2 }{\partial u \partial t'}.
 \end{equation}
\end{subequations}
Using~\eqref{eqa:A-wave_transf} wherever appropriate, the different terms of~\eqref{eqa:A-wave_eq} become
\begin{subequations}\label{eqa:A-new_deriv}
\begin{equation}
c^2 \frac{{{\partial ^2} \mathbf{E}_\text{M}(x,z,t) }}{{\partial {x^2}}}= c^2 \frac{{{\partial ^2} \mathbf{E}_\text{M}(x,z,t) }}{{\partial {x'^2}}},
\end{equation}
\begin{equation}
c^2 \frac{{{\partial ^2} \mathbf{E}_\text{M}(x,z,t) }}{{\partial {z^2}}}= c^2 \beta_\text{m}^2 \frac{{{\partial ^2} \mathbf{E}_\text{M}(x,z,t) }}{{\partial {u^2}}},
\end{equation}
\begin{equation}
\mathbf{E}_\text{M}(x,z,t) \frac{{{\partial ^2} \epsilon (z,t) }}{{\partial {t^2}}}=\mathbf{E}_\text{M}(x,z,t)  \frac{\partial^2 }{\partial t'^2} \left(\sum\limits_{k = - \infty }^\infty  \tilde{\epsilon}_k e^{j k u}\right)=-\omega_\text{m}^2 \mathbf{E}_\text{M}(x,z,t)  \sum\limits_{\substack{k =  - \infty\\k \neq  0}} ^\infty  k^2 \tilde{\epsilon}_ke^{j k u},
\end{equation}
\begin{equation}
\epsilon (z,t) \frac{{{\partial ^2} \mathbf{E}_\text{M}(x,z,t) }}{{\partial {t^2}}}=\sum\limits_{k = - \infty }^\infty  \tilde{\epsilon}_k e^{j k u} \left( \frac{\partial^2 \mathbf{E}_\text{M}(x,z,t)}{\partial t'^2}+\omega_\text{m}^2 \frac{\partial^2 \mathbf{E}_\text{M}(x,z,t) }{\partial u^2}+2 \omega_\text{m} \frac{\partial^2 \mathbf{E}_\text{M}(x,z,t)}{\partial u \partial t'}  \right),
\end{equation}
\begin{equation}
\begin{split}
2\frac{{{\partial } \epsilon (z,t) }}{{\partial {t}}} \frac{{{\partial \mathbf{E}_\text{M}(x,z,t) } }}{{\partial {t}}}&= 2 \omega_\text{m} \frac{\partial }{\partial u} \left(\sum\limits_{k = - \infty }^\infty  \tilde{\epsilon}_k e^{j k u} \right)   \left( \frac{\partial \mathbf{E}_\text{M}(x,z,t) }{\partial t'}+\omega_\text{m} \frac{\partial \mathbf{E}_\text{M}(x,z,t) }{\partial u}  \right) \\&=2 j  \omega_\text{m} \sum\limits_{\substack{k =  - \infty\\k \neq  0} }^\infty  k \tilde{\epsilon}_k e^{j k u}    \left( \frac{\partial \mathbf{E}_\text{M}(x,z,t) }{\partial t'}+\omega_\text{m} \frac{\partial \mathbf{E}_\text{M}(x,z,t) }{\partial u}  \right),
\end{split}
\end{equation}
\end{subequations}

Grouping~\eqref{eqa:A-new_deriv} according to~\eqref{eqa:A-wave_eq} yields then the wave equation in terms of $x',u,t'$ and $\tilde{\epsilon}_k$:
\begin{equation}
\begin{split}
\left(c^2 \beta_\text{m}^2 - \omega_\text{m}^2 \sum\limits_{k = - \infty }^\infty  \tilde{\epsilon}_k e^{j k u}  \right)  \frac{{{\partial ^2} \mathbf{E}_\text{M}(x,z,t) }}{{\partial {u^2}}} +c^2 \frac{{{\partial ^2} \mathbf{E}_\text{M}(x,z,t) }}{{\partial {x'^2}}} -  \sum\limits_{k = - \infty }^\infty  \tilde{\epsilon}_k e^{j k u} \frac{\partial^2 \mathbf{E}_\text{M}(x,z,t)}{\partial t'^2} \\ -2 \omega_\text{m} \sum\limits_{k = - \infty }^\infty  \tilde{\epsilon}_k e^{j k u} \frac{\partial^2 \mathbf{E}_\text{M}(x,z,t)}{\partial u \partial t'}  -2 j \omega_\text{m} \sum\limits_{\substack{k =  - \infty\\k \neq  0}}^\infty  k  \tilde{\epsilon}_k e^{j k u}    \left( \frac{\partial \mathbf{E}_\text{M}(x,z,t) }{\partial t'}+\omega_\text{m} \frac{\partial \mathbf{E}_\text{M}(x,z,t) }{\partial u}  \right) \\+ \omega_\text{m}^2 \sum\limits_{\substack{k =  - \infty\\k \neq  0}} ^\infty  k^2 \tilde{\epsilon}_k e^{j k u} \mathbf{E}_\text{M}(x,z,t)=0.
\end{split}
\label{eqa:A-long_eq_u_tp}
\end{equation}
For this equation to really represent the wave equation, it must maintain all of its order derivatives. This is generally the case, except when the coefficient of the first term vanishes, i.e.
\begin{equation}
c^2 \beta_\text{m}^2 - \omega_\text{m}^2 \tilde{\epsilon}_0 - \omega_\text{m}^2 \sum\limits_{\substack{k =  - \infty\\k \neq  0}}^\infty  \tilde{\epsilon}_k e^{j k u}=0,
\label{eqa:A-ggge}
\end{equation}
or, assuming a real permittivity and hence $\Im\left\{\sum\limits_{k = - \infty }^\infty  \tilde{\epsilon}_k e^{j k u}\right\}=0$,
\begin{equation}
 \sum\limits_{\substack{k =  - \infty\\k \neq  0}}^\infty  \tilde{\epsilon}_k e^{j k u}= \frac{\epsilon_\text{r} }{\gamma^2}   -  \tilde{\epsilon}_0,
\label{eqa:A-summ_sonic}
\end{equation}
\noindent where~\eqref{eqa:vm_omegam_betam}, \eqref{eqa:vr} and~\eqref{eqa:gamma} have been used. Assuming that the permittivity variation is bounded to $\epsilon_\text{m}$, i.e. \mbox{$\bigg|\sum\limits_{\substack{k =  - \infty\\k \neq  0}}^\infty  \tilde{\epsilon}_k e^{j k u}\bigg| \leq \epsilon_\text{m}$}, and considering that $u$ is real, the condition~\eqref{eqa:A-summ_sonic} reduces to
\begin{equation}
\left| \frac{\epsilon_\text{r} }{\gamma^2}   -  \tilde{\epsilon}_0\right| \leq \epsilon_\text{m},
\label{eqa:A-son_cond}
\end{equation}
which may be rearranged as
\begin{equation}\label{eqa:A-sonic_sin}
\sqrt{\frac{\epsilon_\text{r}}{\tilde{\epsilon}_0+\epsilon_\text{m}}}
\leq\gamma\leq
\sqrt{\frac{\epsilon_\text{r}}{\tilde{\epsilon}_0-\epsilon_\text{m}}}.
\end{equation}
This corresponds to the sonic interval, where the solution for $\mathbf{E}_\text{m}$ cannot be computed using the Bloch-Floquet expansion~\eqref{eqa:wave}.

\subsection{General Wave Solution}\label{sec:A-Gen}
To solve the wave equation in~\eqref{eqa:wave_eq}, we write the product of~\eqref{eqa:Fourier_perm} and~\eqref{eqa:wave} as
\begin{equation}
\begin{split}
\epsilon (z,t)\mathbf{E}_\text{M}(x,z,t) = \mathbf{\hat{y}} \sum\limits_{n =-\infty }^\infty  \sum\limits_{k =- \infty }^\infty  \tilde{\epsilon}_k   A_{n}^\pm e^{-j ( \pm \beta_{0} z +k_x x-\omega_0 t )}    e^{  -j(n+k)( \beta_\text{m} z-\omega _\text{m}t )} \\
= \sum\limits_{n =-\infty }^\infty  \sum\limits_{k =- \infty }^\infty  \tilde{\epsilon}_k   A_{n-k}^\pm  e^{-j (\pm \beta_{0} z +k_x x- \omega_0 t )}   e^{  -jn(\beta_\text{m} z-\omega _\text{m}t )}.
\end{split}
\label{eqa:A-product_E_eps}
\end{equation}

Inserting~\eqref{eqa:wave} and~\eqref{eqa:A-product_E_eps} into~\eqref{eqa:wave_eq} yields
\begin{equation}
\left(\frac{{{\partial ^2}}}{{\partial {x^2}}} +\frac{{{\partial ^2}}}{{\partial {z^2}}} \right)\sum\limits_{n=  - \infty }^\infty  A_{n}^\pm    e^{  -j(\pm \beta_0z + n\beta_\text{m}z+k_x x)}e^{j (\omega_0 + n\omega_\text{m})t } - \frac{1}{{{c^2}}}\frac{{{\partial ^2} }}{{\partial {t^2}}} \sum\limits_{n =-\infty }^\infty  \sum\limits_{k =- \infty }^\infty  \tilde{\epsilon}_k   A_{n-k}^\pm  e^{-j ( \pm \beta_{0} z +k_x x -\omega_0 t )}   e^{  -jn(\beta_\text{m} z-\omega _\text{m}t )}=0.
\label{eqa:A-deriv_wave_eq}
\end{equation}
Applying the second derivatives transforms this equation to
\begin{equation}\label{eqa:A-summation_A}
\begin{split}
\sum\limits_{n=  - \infty }^\infty  \bigg[ \left(  -k_x^2- (\beta_{0} \pm n\beta_\text{m})^2 \right) & A_n^\pm e^{  -j(\pm \beta_0z + n\beta_\text{m}z+k_x x)} e^{j (\omega_0 + n\omega_\text{m})t}       \\
&+\frac{(\omega_0+n\omega _\text{m} )^2}{c^2}  \sum\limits_{k =  - \infty }^\infty  \tilde{\epsilon}_k A_{n-k}^\pm e^{  -j(\pm \beta_0 + n\beta_\text{m}z+k_x x)} e^{j(\omega_0+ n\omega_\text{m}t)}   \bigg] =0.
\end{split}
\end{equation}
Finally, using the orthogonality property of the complex exponential function to cancel the common terms $e^{  -j(\pm \beta_0z + n\beta_\text{m}z+k_x x)} e^{j(\omega_0+ n\omega_\text{m}t)}$ leads to the recursive set of equations~\eqref{eqa:recurs_gen}.

\subsection{Application of Boundary Conditions}\label{A-BCs}
\subsubsection{Forward Problem}\label{A-BCs_forward}
The TM$_{xz}$ or $E_y$ incident fields read
\begin{subequations}
\begin{equation}
\mathbf{E}_\text{I}^\text{F} (x,z,t)= \mathbf{\hat{y}} E_0 e^{-j\left[k_0\sin(\theta_\text{i}) x +k_0 \cos(\theta_\text{i}) z- \omega _0 t  \right]},
\end{equation}
where $E_0$ is the amplitude of the incident field and $k_0=\omega _0/v_\text{b}$, and
\begin{equation}
\mathbf{H}_\text{I}^\text{F} (x,z,t)= \frac{1}{\eta_1}\left[\mathbf{\hat{k}}_\text{I}^\text{F}  \times \mathbf{E}_\text{I}^\text{F} (x,z,t)\right] =  \left[-\mathbf{\hat{x}} \cos(\theta_\text{i}) +\mathbf{\hat{z}} \sin(\theta_\text{i}) \right] \sqrt{\frac{\epsilon_0 \epsilon_\text{r}}{\mu_0}} E_0 e^{-j\left[k_0\sin(\theta_\text{i}) x +k_0 \cos(\theta_\text{i}) z- \omega _0 t  \right]},
\end{equation}
\end{subequations}
\noindent where $\eta_1=\sqrt{\mu_0/(\epsilon_0\epsilon_r)}$. The electric and magnetic fields in the slab may be explicitly written using~\eqref{eqa:wave} as
\begin{subequations}
\begin{equation}
\mathbf{E}_\text{M}^\text{F}(x,z,t)=\mathbf{\hat{y}}\sum_{n,p =  - \infty}^\infty    \left( A_{np}^{\text{F}+}  e^{ - j \left[ k_x x+ (\beta_{0p}^{+} + n \beta_\text{m})z \right]} + A_{np}^{\text{F}-} e^{-j\left[(k_x x-(\beta_{0p}^{-} - n\beta_\text{m})z \right]} \right) e^{j ( \omega_0 + n\omega _\text{m})t},
\label{eqa:A-E_mod_field}
\end{equation}
and
\begin{equation}
\begin{split}
\mathbf{H}_\text{M}^\text{F}(x,z,t)=&
\frac{1}{\eta_2}\left[\mathbf{\hat{k}}_\text{M}^\text{F}  \times \mathbf{E}_\text{M}^\text{F} (x,z,t)\right]\\
&=\sum_{n,p =  - \infty}^\infty     \bigg( \bigg[-\mathbf{\hat{x}} \frac{\beta_{0p}^{+} + n \beta_\text{m}}{\mu_0 (\omega_0 + n\omega_\text{m})} + \mathbf{\hat{z}} \sin(\theta_\text{i})\sqrt{\frac{\epsilon_0 \epsilon_\text{r}}{\mu_0}} \bigg] A_{np}^{\text{F}+}  e^{ - j \big[k_x x+ (\beta_{0p}^{+} + n \beta_\text{m})z \big]} \\
&+ \bigg[\mathbf{\hat{x}}\frac{\beta_{0p}^{-} - n \beta_\text{m}}{\mu_0 (\omega_0 + n\omega_\text{m})} + \mathbf{\hat{z}} \sin(\theta_\text{i})\sqrt{\frac{\epsilon_0 \epsilon_\text{r}}{\mu_0}} \bigg] A_{np}^{\text{F}-} e^{-j\big[k_x x-(\beta_{0p}^{-} - n\beta_\text{m})z \big]} \bigg)e^{j ( \omega_0 + n\omega_\text{m})t}.
\end{split}
\label{eqa:A-H_mod_field}
\end{equation}
\end{subequations}

\noindent where $\eta_2=\sqrt{\mu_0/(\epsilon_0\epsilon_r)}=\eta_1$. The reflected and transmitted electric fields outside of the slab may be defined as
\begin{subequations}\label{eqa:A-ER_ET_forw}
\begin{equation}
\mathbf{E}_\text{R}^\text{F} (x,z,t)= \mathbf{\hat{y}} \sum\limits_{n =  - \infty }^\infty  E_{\text{r}n}^\text{F} e^{-j \left[ k_{0} \sin(\theta_{\text{i}}) x -k_{0n} \cos(\theta_{\text{r}n}) z-(\omega _0 +n\omega _\text{m}) t    \right] },
\label{eqa:A-E_refl_forw}
\end{equation}
\begin{equation}
\mathbf{H}_\text{R}^\text{F} (x,z,t)= \frac{1}{\eta_1}[\mathbf{\hat{k}}_\text{R}^\text{F}  \times \mathbf{E}_\text{R}^\text{F} (x,z,t)] = \left[\mathbf{\hat{x}} \cos(\theta_{\text{r}n}) +\mathbf{\hat{z}} \sin(\theta_\text{i}) \right] \sqrt{\frac{\epsilon_0 \epsilon_\text{r}}{\mu_0}} E_{\text{r}n}^\text{F}   e^{-j \left[ k_{0} \sin(\theta_{\text{i}}) x -k_{0n} \cos(\theta_{\text{r}n}) z-(\omega _0 +n\omega _\text{m}) t    \right] },
\label{eqa:A-H_refl_forw}
\end{equation}
\begin{equation}
\mathbf{E}_\text{T}^\text{F} (x,z,t)= \mathbf{\hat{y}} \sum\limits_{n =  - \infty }^\infty  E_{\text{t}n}^\text{F} e^{-j \left[ k_{0} \sin(\theta_{\text{i}}) x +k_{0n} \cos(\theta_{\text{t}n}) z-(\omega _0 +n\omega _\text{m}) t    \right] },
\label{eqa:A-E_trans_forw}
\end{equation}
\begin{equation}
\mathbf{H}_\text{T}^\text{F} (x,z,t)= \frac{1}{\eta_3}[\mathbf{\hat{k}}_\text{T}^\text{F}  \times \mathbf{E}_\text{T}^\text{F} (x,z,t)] = \left[-\mathbf{\hat{x}} \cos(\theta_{\text{t}n}) +\mathbf{\hat{z}} \sin(\theta_\text{i}) \right] \sqrt{\frac{\epsilon_0 \epsilon_\text{r}}{\mu_0}} E_{\text{t}n}^\text{F} e^{-j \left[ k_{0} \sin(\theta_{\text{i}}) x +k_{0n} \cos(\theta_{\text{t}n}) z-(\omega _0 +n\omega _\text{m}) t    \right] },
\label{eqa:A-H_trans_forw}
\end{equation}
\end{subequations}
\noindent where $\eta_3=\sqrt{\mu_0/(\epsilon_0\epsilon_r)}=\eta_1=\eta_2$. We then enforce the continuity of the tangential components of the electromagnetic fields at $z=0$ and $z=L$ to find the unknown field amplitudes $A_{0p}^\pm$, $E_{\text{r}n}^\text{F}$ and $E_{\text{t}n}^\text{F}$. The electric field continuity condition between regions 1 and 2 at $z=0$, ${E_{1y}}(x,0,t) = {E_{2y}}(x,0,t)$, reduces to
\begin{equation}\label{eqa:EBC_forw_12}
\delta_{n0} E_0  + E_{\text{r}n}^\text{F} =  \sum_{p =  - \infty}^\infty  \left( A_{np}^{\text{F}+}+ A_{np}^{\text{F}-} \right),
\end{equation}
\noindent while the corresponding magnetic field continuity condition, ${H_{1x}}(x,0,t) = {H_{2x}}(x,0,t)$, reads
\begin{equation}\label{eqa:HBC_forw_12}
\sqrt{\frac{\epsilon_0 \epsilon_\text{r}}{\mu_0}} \cos(\theta_\text{i})   \delta_{n0} E_0 - \sqrt{\frac{\epsilon_0 \epsilon_\text{r}}{\mu_0}} \cos(\theta_{\text{r}n}) E_{\text{r}n}^\text{F}  =  \sum_{p =  - \infty}^\infty  \left( \frac{\beta_{0p}^{+} + n \beta_\text{m}}{\mu_0 (\omega_0 + n\omega_\text{m})}  A_{np}^{\text{F}+}  - \frac{\beta_{0p}^{-} - n \beta_\text{m}}{\mu_0 (\omega_0 + n\omega_\text{m})}  A_{np}^{\text{F}-}  \right),
\end{equation}
\noindent  where $k_n=(\omega _0 +n\omega _\text{m})/v_\text{b}$.

Similarly, the electric field continuity condition between regions 2 and 3 at $z=L$, ${E_{2y}}(x,L,t) = {E_{3y}}(x,L,t)$, reduces to
\begin{equation}
 \sum_{p =  - \infty}^\infty \left( A_{np}^{\text{F}+} e^{-j (\beta_{0p}^+ + n \beta_\text{m})L} +A_{np}^{\text{F}-} e^{j (\beta_{0p}^- - n \beta_\text{m})L} \right)  = E_{\text{t}n}^\text{F}  e^{- j k_{0n} \cos(\theta_{\text{t}n}) L},
 \label{eqa:EBC_forw_23}
\end{equation}
while the corresponding magnetic field continuity condition between regions 2 and 3 at $z=L$, ${H_{2x}}(x,L,t) = {H_{3x}}(x,L,t)$, reads
\begin{equation}
\sum_{p =  - \infty}^\infty  \left[ \frac{\beta_{0p}^{+} + n \beta_\text{m}}{\mu_0 (\omega_0 + n\omega_\text{m})}  A_{np}^{\text{F}+} e^{-j (\beta_{0p}^+ + n \beta_\text{m})L}  - \frac{\beta_{0p}^{-} - n \beta_\text{m}}{\mu_0 (\omega_0 + n\omega_\text{m})}  A_{np}^{\text{F}-}  e^{j (\beta_{0p}^- - n \beta_\text{m})L} \right]  = \sqrt{\frac{\epsilon_0 \epsilon_\text{r}}{\mu_0}} \cos(\theta_{\text{t}n}) E_{\text{t}n}^\text{F} e^{- j k_{0n} \cos(\theta_{\text{t}n}) L}
 \label{eqa:HBC_forw_23}
\end{equation}
Solving~\eqref{eqa:EBC_forw_23} and~\eqref{eqa:HBC_forw_23}  for $A_{0p}^{\text{F}-}$ yields~\eqref{eqa:R_forward}. Next solving~\eqref{eqa:EBC_forw_12}, \eqref{eqa:HBC_forw_12} and ~\eqref{eqa:R_forward} for $A_{0p}^{\text{F}+}$ yields~\eqref{eqa:T_forward}. Finally, the total scattered fields outside the slab are obtained by substituting~\eqref{eqa:T_forward_backward} into~\eqref{eqa:A-ER_ET_forw}, which yields~\eqref{eqa:E_RT_forward}.
\subsubsection{Backward Problem}\label{A-BCs_backward}
The TM$_{xz}$ or $E_y$ incident fields read
\begin{subequations}
\begin{equation}
\mathbf{E}_\text{I}^\text{B} (x,z,t)= \mathbf{\hat{y}} E_0 e^{-j\left[k_0\sin(\theta_\text{i}) x -k_0 \cos(\theta_\text{i}) z- \omega _0 t  \right]},
\end{equation}
\begin{equation}
\mathbf{H}_\text{I}^\text{B} (x,z,t)= \frac{1}{\eta_1}[\mathbf{\hat{k}}_\text{I}^\text{B}  \times \mathbf{E}_\text{I}^\text{B} (x,z,t)] =  \left[\mathbf{\hat{x}} \cos(\theta_\text{i}) +\mathbf{\hat{z}} \sin(\theta_\text{i}) \right] \sqrt{\frac{\epsilon_0 \epsilon_\text{r}}{\mu_0}} E_0 e^{-j\left[k_0\sin(\theta_\text{i}) x +k_0 \cos(\theta_\text{i}) z- \omega _0 t  \right]},
\end{equation}
\end{subequations}
The electric and magnetic fields in the slab may be explicitly written using~\eqref{eqa:wave} as
\begin{subequations}
\begin{equation}
\mathbf{E}_\text{M}^\text{B}(x,z,t)=\mathbf{\hat{y}} \sum_{n,p =  - \infty}^\infty     \left( A_{np}^{\text{B}+}  e^{ - j \left[ k_x x+(\beta_{0p}^{+} + n \beta_\text{m})z \right]} + A_{np}^{\text{B}-} e^{- j\left[k_x x-(\beta_{0p}^{-} - n\beta_\text{m})z \right]} \right) e^{j ( \omega_0 + n\omega _\text{m})t},
\label{eqa:A-E_mod_field_back}
\end{equation}
\begin{equation}
\begin{split}
\mathbf{H}_\text{M}^\text{B}(x,z,t)=&
\frac{1}{\eta_2}\left[\mathbf{\hat{k}}_\text{M}^\text{B}  \times \mathbf{E}_\text{M}^\text{B} (x,z,t)\right]\\
&=\sum_{n,p =  - \infty}^\infty     \bigg( \bigg[-\mathbf{\hat{x}} \frac{\beta_{0p}^{+} + n \beta_\text{m}}{\mu_0 (\omega_0 + n\omega_\text{m})} + \mathbf{\hat{z}} \sin(\theta_\text{i})\sqrt{\frac{\epsilon_0 \epsilon_\text{r}}{\mu_0}} \bigg] A_{np}^{\text{B}+}  e^{ - j \big[k_x x+ (\beta_{0p}^{+} + n \beta_\text{m})z \big]} \\
&+ \bigg[\mathbf{\hat{x}}\frac{\beta_{0p}^{-} - n \beta_\text{m}}{\mu_0 (\omega_0 + n\omega_\text{m})} + \mathbf{\hat{z}} \sin(\theta_\text{i})\sqrt{\frac{\epsilon_0 \epsilon_\text{r}}{\mu_0}} \bigg] A_{np}^{\text{B}-} e^{-j\big[k_x x-(\beta_{0p}^{-} - n\beta_\text{m})z \big]} \bigg)e^{j ( \omega_0 + n\omega_\text{m})t}.
\end{split}
\label{eqa:A-H_mod_field_back}
\end{equation}
\end{subequations}
The reflected and transmitted electric fields outside of the slab may be defined as
\begin{subequations}\label{eqa:A-ER_ET_backw}
\begin{equation}
\mathbf{E}_\text{R}^\text{B} (x,z,t)= \mathbf{\hat{y}} \sum\limits_{n =  - \infty }^\infty  E_{\text{r}n}^\text{B} e^{j \left[ (\omega _0 +n\omega _\text{m}) t - k_{0n} \cos(\theta_{\text{r}n}) z- k_{0} \sin(\theta_{\text{i}}) x \right] },
\label{eqa:A-E_refl_backw}
\end{equation}
\begin{equation}
\mathbf{H}_\text{R}^\text{B} (x,z,t)= \frac{1}{\eta_3}[\mathbf{\hat{k}}_\text{R}^\text{B}  \times \mathbf{E}_\text{R}^\text{B} (x,z,t)] = \left[-\mathbf{\hat{x}} \cos(\theta_{\text{r}n}) +\mathbf{\hat{z}} \sin(\theta_\text{i}) \right] \sqrt{\frac{\epsilon_0 \epsilon_\text{r}}{\mu_0}} E_{\text{r}n}^\text{B}    e^{j \left[ (\omega _0 +n\omega _\text{m}) t - k_{0n} \cos(\theta_{\text{r}n}) z- k_{0} \sin(\theta_{\text{i}}) x \right] },
\label{eqa:A-H_refl_backw}
\end{equation}
\begin{equation}
\mathbf{E}_\text{T}^\text{B} (x,z,t)= \mathbf{\hat{y}} \sum\limits_{n =  - \infty }^\infty  E_{\text{t}n}^\text{B} e^{-j \left[ k_{0} \sin(\theta_\text{i}) x -k_{0n} \cos(\theta_{\text{t}n}) z -(\omega _0 +n\omega_\text{m}) t  \right] },
\label{eqa:A-E_trans_backw}
\end{equation}
\begin{equation}
\mathbf{H}_\text{T}^\text{B} (x,z,t)= \frac{1}{\eta_1}[\mathbf{\hat{k}}_\text{T}^\text{B}  \times \mathbf{E}_\text{T}^\text{B} (x,z,t)] = \left[\mathbf{\hat{x}} \cos(\theta_{\text{t}n}) +\mathbf{\hat{z}} \sin(\theta_\text{i}) \right] \sqrt{\frac{\epsilon_0 \epsilon_\text{r}}{\mu_0}} E_{\text{t}n}^\text{B} e^{-j \left[ k_{0} \sin(\theta_\text{i}) x -k_{0n} \cos(\theta_{\text{t}n}) z -(\omega _0 +n\omega_\text{m}) t  \right] }.
\label{eqa:A-H_trans_backw}
\end{equation}
\end{subequations}
 We then enforce the continuity of the tangential components of the electromagnetic fields at $z=0$ and $z=L$ to find the unknown field amplitudes $A_{0p}^\pm$, $E_{\text{r}n}^\text{B}$ and $E_{\text{t}n}^\text{B}$. The electric field continuity condition between regions 2 and 3 at $z=L$, $E_{2y}(x,L,t) = E_{3y}(x,L,t)$, reduces to
\begin{equation}\label{eqa:EBC_backw_23}
{E_0}{\delta _{n0}} e^{j k_\text{0} \cos(\theta_\text{i}) L}+ E_{\text{r}n}^\text{B} {e^{ - j k_{0n} \cos(\theta_{\text{r}n}) L}}=  \sum_{p =  - \infty}^\infty \left( A_{np}^{\text{B}+} e^{-j (\beta_{0p}^+ + n \beta_\text{m})L} +A_{np}^{\text{B}-} e^{j (\beta_{0p}^- - n \beta_\text{m})L} \right),
\end{equation}
\noindent while the corresponding magnetic field continuity condition, ${H_{2x}}(x,0,t) = {H_{3x}}(x,0,t)$, reads
\begin{equation}\label{eqa:HBC_backw_23}
\begin{split}
\sqrt{\frac{\epsilon_0 \epsilon_\text{r}}{\mu_0}} \cos(\theta_\text{i}) &  \delta_{n0} E_0 e^{j k_\text{0} \cos(\theta_\text{i}) L} - \sqrt{\frac{\epsilon_0 \epsilon_\text{r}}{\mu_0}} \cos(\theta_{\text{r}n}) E_{\text{r}n}^\text{B} {e^{ - j k_{0n} \cos(\theta_{\text{r}n}) L}} \\
& =  \sum_{p =  - \infty}^\infty  \left( - \frac{\beta_{0p}^{+} + n \beta_\text{m}}{\mu_0 (\omega_0 + n\omega_\text{m})}  A_{np}^{\text{B}+}  e^{-j (\beta_{0p}^+ + n \beta_\text{m})L} + \frac{\beta_{0p}^{-} - n \beta_\text{m}}{\mu_0 (\omega_0 + n\omega_\text{m})}  A_{np}^{\text{B}-} e^{j (\beta_{0p}^- - n \beta_\text{m})L} \right).
\end{split}
\end{equation}
Similarly, the electric field continuity condition between regions 1 and 2 at $z=0$, ${E_{1y}}(x,0,t) = {E_{2y}}(x,0,t)$, reduces to
\begin{equation}\label{eqa:EBC_backw_12}
\sum_{p =  - \infty}^\infty  \left(  A_{np}^{\text{B}+} +A_{np}^{\text{B}-}\right)  = E_{\text{t}n}^\text{B}
\end{equation}
while the corresponding magnetic field continuity condition between regions 1 and 2 at $z=0$, ${H_{1x}}(x,0,t) = {H_{2x}}(x,0,t)$, reads
\begin{equation}\label{eqa:HBC_backw_12}
\sum_{p =  - \infty}^\infty  \left[ -\frac{\beta_{0p}^{+} + n \beta_\text{m}}{\mu_0 (\omega_0 + n\omega_\text{m})}  A_{np}^{\text{B}+}  + \frac{\beta_{0p}^{-} - n \beta_\text{m}}{\mu_0 (\omega_0 + n\omega_\text{m})}  A_{np}^{\text{B}-}   \right]  = \sqrt{\frac{\epsilon_0 \epsilon_\text{r}}{\mu_0}} \cos(\theta_{\text{t}n}) E_{\text{t}n}^\text{B}
\end{equation}
Solving~\eqref{eqa:EBC_backw_12} and~\eqref{eqa:HBC_backw_12}  for $A_{0p}^{\text{B}+}$ yields~\eqref{eqa:R_backward}. Next solving~\eqref{eqa:EBC_backw_23}, \eqref{eqa:HBC_backw_23} and~\eqref{eqa:R_backward} for $A_{0p}^{\text{B}-}$ yields~\eqref{eqa:T_backward}. Finally, the total scattered fields outside the slab are obtained by substituting~\eqref{eqa:T_forward_backward_b} into~\eqref{eqa:A-ER_ET_backw}, which yields~\eqref{eqa:E_RT_backward}.

\end{document}